\documentclass[aps,prd,groupaddress,tighten,nofootinbib,preprint]{revtex4}
\usepackage{amssymb}
\usepackage{amsmath}
\usepackage{color}
\usepackage{epsfig}
\usepackage{graphicx,epsf,epsfig}
\usepackage{bbm}
\usepackage{mciteplus}
\def\slc#1{\setbox0=\hbox{$#1$}           % set a box for #1
    \dimen0=\wd0                                 % and get its size
    \setbox1=\hbox{/} \dimen1=\wd1               % get size of /
    \ifdim\dimen0>\dimen1                        % #1 is bigger
       \rlap{\hbox to \dimen0{\hfil/\hfil}}      % so center / in box
       #1                                        % and print #1
    \else                                        % / is bigger
       \rlap{\hbox to \dimen1{\hfil$#1$\hfil}}   % so center #1
       /                                         % and print /
    \fi}

\begin{document}
%-----------------------------------------------------------------------------------------
\title{Renormalization group running of neutrino parameters\\ in the inverse seesaw model}
%-----------------------------------------------------------------------------------------
%\date{\today}
%-----------------------------------------------------------------------------------------

\author{Johannes Bergstr{\"o}m}
\email{johbergs@kth.se}

\author{Michal Malinsk\'y}
\email{malinsky@kth.se}

\author{Tommy Ohlsson}
\email{tommy@theophys.kth.se}

\author{He Zhang}
\email{zhanghe@kth.se}

\affiliation{Department of Theoretical Physics, School of
Engineering Sciences, Royal Institute of Technology (KTH) --
AlbaNova University Center, Roslagstullsbacken 21, 106 91 Stockholm,
Sweden}

\begin{abstract}
We perform a detailed study of the renormalization group equations
in the inverse seesaw model. Especially, we derive compact
analytical formulas for the running of the neutrino parameters in
the standard model and the minimal supersymmetric standard model,
and illustrate that, due to large Yukawa
coupling corrections, significant running effects on the leptonic
mixing angles can be naturally obtained in the proximity of the
electroweak scale, perhaps even within the reach of the LHC. In
general, if the mass spectrum of the light neutrinos is nearly
degenerate, the running effects are enhanced to experimentally
accessible levels, well suitable for the investigation of the
underlying dynamics behind the neutrino mass generation and the
lepton flavor structure. In addition, the effects of the seesaw
thresholds are discussed, and a brief comparison to other seesaw
models is carried out.
\end{abstract}


\maketitle

\section{Introduction}
\label{sec:introduction}

Experimental progress on neutrino masses and leptonic mixing has
opened up a new window in searching for new physics beyond the
standard model (SM) of particle physics during the past decade.
Since neutrinos are massless particles in the SM, one usually
extends the SM particle content in order to accommodate massive
neutrinos.  Among various theories of this kind, the seesaw
mechanism~\cite{Minkowski:1977sc,*Yanagida:1979as,*GellMann:1980vs,*Mohapatra:1979ia,*Magg:1980ut,*Schechter:1980gr,*Wetterich:1981bx,*Lazarides:1980nt,*Mohapatra:1980yp,*Cheng:1980qt,*Foot:1988aq}
attracts a lot of attention in virtue of its naturalness and
simplicity. For instance, in the conventional type-I seesaw model,
three right-handed neutrinos with a Majorana mass term far above the
electroweak scale are introduced. The masses of the light neutrinos
are then strongly suppressed with respect to the masses of the
charged leptons by the ratio between the electroweak scale and the
mass scale of the heavy right-handed neutrinos. Usually, the
neutrino parameters are measured in low-energy scale experiments,
while on the other hand, the seesaw-induced neutrino mass operator
often emerges at some very high-energy scale. Therefore, neutrino
masses and leptonic mixing parameters are subject to radiative
corrections, i.e., they are modified by the renormalization group
(RG) running. In principle, the RG running effects can even be
physically relevant, especially if the seesaw scale turns out to be
not extremely high.

The generic features of the RG running of neutrino parameters have
been investigated intensively in the literature. Typically, at
energy scales lower than the {\it seesaw threshold}, i.e., the mass
scale of the heavy seesaw particles, the RG running behavior of
neutrino masses and leptonic mixing can be described within an
effective theory, which is essentially the same for various seesaw
models, reflecting the uniqueness of the dimension-five Majorana
mass operator in the SM. However, at energy scales higher than the
seesaw threshold, a full theory has to be considered, and the
interplay between the light and heavy sectors could make the RG
running effects particularly different compared to those in the
effective theory. The RG running effects above the seesaw scale are
especially relevant in the wide class of popular theories based on
the idea of grand unification, i.e. grand unification theory (GUT),
where the specific flavor structure stipulated at the GUT scale,
typically of the order of $10^{16}$~GeV, often experiences a
long range running over many intermediate energy-scale thresholds.

Concerning any specific seesaw model, a generic analysis of the RG
running of neutrino parameters is clearly inconceivable because of
an infinite number of possible underlying models, in particular,
above the relevant seesaw threshold. Nevertheless, the universal
features, like the presence of extra degrees of freedom underpinning
the seesaw, i.e., the heavy fermions in the type-I and -III models
or the scalar triplets in the type-II seesaw model, and the nature
of their interactions with the SM sector, still admit for a high
degree of theoretical scrutiny, providing a basis for any further
work within a specific class of unified models.

The full set of renormalization group equations (RGEs) in the type-I, -II, and -III seesaw models
have been derived, both in the SM and in the minimal supersymmetric
standard model
(MSSM)~\cite{Chankowski:1993tx,*Babu:1993qv,*Antusch:2001ck,*Antusch:2001vn,Rossi:2002zb,*Joaquim:2009vp,Chao:2006ye,*Schmidt:2007nq,Chakrabortty:2008zh}.
However, in these models, the seesaw scale is often taken not too
far below the GUT scale, which hinders the direct experimental
testability of the origin of neutrino masses. Although, in the
type-I seesaw model, one can bring down the mass scale of the
right-handed neutrinos by means of a severe fine-tuning among the
model
parameters~\cite{Pilaftsis:1991ug,*Kersten:2007vk,*Antusch:2009gn,*Zhang:2009ac},
radiative corrections induced by right-handed neutrinos tend to
spoil the stability of such settings beyond the tree-level
approximation.

Recently, a lot of attention has been focused on low-scale seesaw
models, and especially, the possibility of searching for seesaw
particles at the Large Hadron Collider (LHC), see,
e.g., Ref.~\cite{Nath:2010zj} and references therein. In these
models, the smallness of the neutrino masses is protected by other
means than the GUT-scale suppression, such as a small amount of
lepton number breaking. A very attractive example along this
direction is the inverse seesaw model~\cite{Mohapatra:1986bd}, in
which the light neutrino masses are driven by a tiny Majorana mass
insertion in the heavy Dirac neutrino mass matrix instead of the
mass scale of the heavy neutrinos. Furthermore, due to the
pseudo-Dirac feature of the heavy singlets, the model does not
suffer from either large radiative corrections or unnatural
fine-tuning problems. This admits for bringing the heavy neutrinos
down to the LHC energy range while retaining potentially large
Yukawa couplings. Subsequently, this makes the model
phenomenologically very attractive from the lepton flavor violation
point of view~\cite{Bernabeu:1987gr} or for the potentially large
nonunitarity effects in the leptonic mixing
matrix~\cite{Antusch:2006vwa}. Since the RG running, and in
particular, the threshold effects, in such a scenario can play an
important role even at a relatively low energy scale, there is a
need to look in detail at the running effects on neutrino
parameters.

In this work, we will investigate in detail the RG evolution of
neutrino masses and leptonic mixing parameters in the inverse seesaw
model. In particular, in Sec.~\ref{sec:formula}, we first review
briefly the inverse seesaw model. Next, in Sec.~\ref{sec:RGE}, we
present the full set of RGEs for the neutrino parameters. Then, in
Sec.~\ref{sec:approxRGEs}, we provide useful approximations of the
RGEs found in Sec.~\ref{sec:RGE}. Section~\ref{sec:numeric} is
devoted to detailed numerical illustrations and interpretation of
the RG running behavior of the light neutrino mass and leptonic
mixing parameters. In Sec.~\ref{sec:comparison}, a discussion and
comparisons between RGEs in various seesaw models are performed.
Finally, in Sec.~\ref{sec:summary}, a summary is given and our
conclusions are presented. In addition, in App.~\ref{app:exactRGEs},
the complete one-loop RGEs for some of the neutrino parameters are
listed.

\section{The inverse seesaw model}
\label{sec:formula}

The inverse seesaw model is constructed by extending the SM particle
content with three right-handed neutrinos $\nu_{\rm R} = (\nu_{\rm
  R1},\nu_{\rm R2},\nu_{\rm R3})$ and three SM gauge singlets
$S=(S_1,S_2,S_3)$.\footnote{Note that in order to accommodate the experimentally measured neutrino mass-squared differences and mixing angles, the minimal setup is to introduce only two right-handed neutrinos together with two heavy singlets \cite{Malinsky:2009df}, which leads to one massless neutrino in the model. In this work, we will concentrate on the more general case with three right-handed neutrinos together with three heavy singlets, which is the minimal case allowing three massive light neutrinos.}
The Lagrangian of the lepton masses is then
arranged so that it reads in flavor basis
\begin{eqnarray}\label{eq:Lmass}
-{\cal L}_{\nu} = \overline{\ell_{\rm L}} \phi Y_e e_{\rm R} +
\overline{\ell_{\rm L}} \tilde\phi Y_{\nu} \nu_{\rm R} +
\overline{S^c} M_{\rm R} \nu_{\rm R} + \frac{1}{2} \overline{S^c}
M_S S  + {\rm h.c.}\, ,
\end{eqnarray}
where $\phi$ is the SM Higgs fields with $\tilde{\phi} = {\rm
i}\tau_2 \phi^*$, $M_{\rm R}$ is an arbitrary $3\times 3$ matrix,
and $M_S$ is a $3 \times 3$ complex symmetric matrix. Here $Y_e$ and
$Y_\nu$ are the corresponding Yukawa coupling matrices, and in
general, they are arbitrary. Without loss of generality, due to the
freedom in the right-handed fields, one can always perform basis
transformations to keep both $M_{\rm R}$ and $Y_e$ diagonal, i.e.,
$M_{\rm R} = {\rm
  diag} (M_1,M_2,M_3)$ and $Y_{e} = {\rm diag}
(y_e,y_\mu,y_\tau)$. Since the new extra fermions are SM gauge
singlets, their masses are not protected by the Higgs mechanism, and
could be related with some new physics beyond the SM. Thus, one
expects the scale of heavy seesaw particles to be higher than the
electroweak scale, i.e., $M_{\rm R} > {\cal O}(100)~{\rm GeV}$. At
energy scales lower than $M_{\rm R}$, heavy degrees of freedom in the
theory should be integrated out, which leaves a series of
higher-dimensional nonrenormalizable operators in the effective
theory. At dimension-five level, the only allowed operator is the
so-called Weinberg operator, coupling two lepton doublets to the SM
Higgs field, on the form
\begin{eqnarray}\label{eq:operator}
{\cal L}^{d=5}_{\nu} = -\frac{1}{2} (\overline{\ell_{\rm L}} \phi)
\cdot \kappa \cdot (\phi^T \ell^c_{\rm L}) + {\rm h.c.} \, ,
\end{eqnarray}
where $\kappa$ is the effective coupling matrix. In the inverse seesaw
model and at tree-level, it is given by
\begin{eqnarray}\label{eq:kappa}
\kappa \simeq Y_\nu M^{-1}_{\rm R} M_S (M^{T}_{\rm R})^{-1} Y^T_\nu
\, .
\end{eqnarray}
After electroweak symmetry breaking, the dimension-five operator
defined in Eq.~\eqref{eq:operator} yields an effective Majorana mass
term for the light neutrinos
\begin{eqnarray}\label{eq:Mnu}
m_\nu = \kappa v^2 \, ,
\end{eqnarray}
with $v \simeq 174~{\rm GeV}$ being the vacuum expectation value of
the Higgs doublet.

Comparing Eq.~\eqref{eq:kappa} with the typical type-I seesaw model,
in which the effective coupling matrix $\kappa$ is given by $\kappa
\simeq Y_\nu M^{-1}_{\rm R} Y^T_\nu$, one can observe that, in the
inverse seesaw, the masses of the light neutrinos are not only
suppressed by $M_{\rm R}$, but also by the small Majorana insertion
$M_S$. In the limit $M_S \to 0$, $\kappa=0$ indicates the restoration
of lepton number conservation, which in turn ensures the naturalness
of a tiny $M_S$.

Now, since the energy scale of $M_{\rm R}$ is not necessarily high,
we may naturally ask if it is related to some new physics around the
TeV scale and could be tested at the current hadron colliders (e.g., LHC). In fact, for the purpose of searching for heavy singlets at
colliders, it is very important to have visible mixing effects
between the light and heavy neutrinos. Furthermore, the admixture
results in nonunitarity effects in neutrino flavor transitions,
especially for some of the future long-baseline neutrino oscillation
experiments~\cite{Antusch:2006vwa,Antusch:2008tz,Malinsky:2009gw,Malinsky:2009df,FernandezMartinez:2007ms,*Abada:2007ux,*Xing:2007zj,*Goswami:2008mi,*Luo:2008vp,*Altarelli:2008yr,*Antusch:2009pm,*Rodejohann:2009cq}.

In spite of the underlying physics responsible for large neutrino
masses, the particle content of the inverse seesaw model is
essentially the same as that of the type-I seesaw model, but with
six right-handed neutrinos. In principle, one may treat the heavy
singlets $S_i$ ($i=1,2,3$) as three extra right-handed neutrinos,
possessing vanishing Yukawa couplings with the lepton doublets. In
the limit $M_S \to 0$, lepton number conservation is restored, and
the six heavy singlets can be combined together to form three
four-component Dirac particles, while keeping the light neutrinos
massless. Therefore, any phenomena in association with lepton number
violation should be proportional to the Majorana mass insertion
$M_S$.

\section{RGEs for the neutrino mass matrix}
\label{sec:RGE}

As discussed above, the analogy between the type-I seesaw model and
the inverse seesaw model allows us to readily write down the RGEs
for the Yukawa couplings and the neutrino mass matrix from the
existing RGEs of the type-I seesaw
model~\cite{Chankowski:1993tx,*Babu:1993qv,*Antusch:2001ck,*Antusch:2001vn}.
Especially, the relevant beta functions, obtained using the minimal
subtraction renormalization scheme, are given by
\begin{eqnarray}
16\pi^2 \mu \frac{{\rm d}Y_e}{{\rm d}\mu} & = &  \left( \alpha_e +
C^e_e X_e + C^\nu_e X_\nu\right)Y_e \, , \label{eq:RGE1}
 \\
16\pi^2 \mu \frac{{\rm d}Y_\nu}{{\rm d}\mu} & = &  \left( \alpha_\nu
+ C^e_\nu X_e + C^\nu_\nu X_\nu \right)Y_\nu \, ,
 \\
16\pi^2 \mu \frac{{\rm d}M_{\rm R}}{{\rm d}\mu} & = & C_{\rm R}
M_{\rm R} Y^\dagger_\nu Y_\nu \, ,  \\
16\pi^2 \mu \frac{{\rm d}M_{S}}{{\rm d}\mu} & = &   0 \, , \label{eq:RGE4}
\end{eqnarray}
where $\mu$ is the renormalization scale, $Y_f$ (for $f=u,d,e,\nu$)
denote the Yukawa coupling matrices with $X_f = Y_f Y^\dagger_f$.
The coefficients $(C^e_e,C^\nu_e,C^e_\nu,C^\nu_\nu,C_{\rm R})$ stand
for $(3/2,-3/2,-3/2,3/2,1)$ in the SM and $(3,1,1,3,2)$ in the MSSM,
respectively.\footnote{Recall that this difference is, namely, due to
the fact that the proper vertex corrections are absent above the
soft supersymmetry-breaking scale in the MSSM. The differences in signs turn
out to be very important for the RG behavior of the leptonic mixing
angles above the seesaw threshold, see,
e.g.,~Ref.~\cite{Antusch:2005gp,*Mei:2005qp} and references therein.}
The coefficient $\alpha_\nu$ is flavor blind and reads
\begin{eqnarray}\label{eq:alpha}
\alpha_\nu  =  {\rm tr} \left(3 X_u + 3 X_d + X_e +  X_\nu
\right)-\frac{9}{20} g^2_1 - \frac{9}{4} g^2_2
\end{eqnarray}
in the SM, and
\begin{eqnarray}
\alpha_\nu  = {\rm tr} \left(3 X_u +  X_\nu \right)  -\frac{3}{5}
g^2_1 - 3 g^2_2
\end{eqnarray} in the MSSM with $g_i$ being the gauge couplings. Then, if we make
use of $m_\nu$ at energy scales above the seesaw threshold, we can derive
from Eqs.~\eqref{eq:kappa} and \eqref{eq:RGE1}-\eqref{eq:RGE4}
\begin{eqnarray}\label{eq:RGEkappa}
\frac{{\rm d}m_\nu}{{\rm d}t} \equiv \dot m_\nu  =   2 \alpha_\nu m_\nu + \left( C^e_\nu X_e + C_m X_\nu \right) m_\nu \\
+   m_\nu \left(
C^e_\nu X_e + C_m X_\nu \right)^T \,,
\end{eqnarray} with $C_m = 1/2$ in
the SM and $C_m=1$ in the MSSM. Here, for simplicity, we have defined
$t=1/(16\pi^2)\ln (\mu/M_{Z})$.

The beta functions in Eqs.~\eqref{eq:RGE1}-\eqref{eq:RGE4} are
obtained in the minimal subtraction renormalization scheme, and thus, we use an
effective theory below the seesaw thresholds, in which the heavy
fields are absent and their physical effects are covered by a series
of higher-dimensional operators.  To ensure that the full and the
effective theories give identical predictions for physical
quantities at low-energy scales, the parameters of the full and the
effective theories have to be related to each other.

In the case of the neutrino mass matrix, this means relations
between the effective coupling matrix $\kappa$ and the parameters
$Y_\nu$, $M_{\rm R}$, and $M_S$ of the full theory. This is
technically called {\it matching} between the full and the effective
theories. In the effective theory, the Weinberg operator
\eqref{eq:operator}, responsible for the masses of the light
neutrinos, does not depend on the specific seesaw realization and
its evolution equation reads
\begin{eqnarray}\label{eq:RGEkappaB}
\dot \kappa  =  \alpha_\kappa \kappa + \left( C^e_\nu X_e  \right)
\kappa + \kappa \left( C^e_\nu X_e \right)^T \, ,
\end{eqnarray}
where
\begin{eqnarray}\label{eq:alphakappa}
\alpha_\kappa & = & 2\:{\rm tr}
\left(3X_u+3X_d+X_e\right) + \lambda - 3g^2_2 \, ,  \qquad \text{(in the SM)}\\
\alpha_\kappa & = & {\rm tr} \left(6X_u\right)  -\frac{6}{5}g^2_1 -
6g^2_2 \, , \qquad \text{(in the MSSM)}
\end{eqnarray}
with $\lambda$ denoting the SM Higgs self-coupling constant.

For the simplest case, if the mass spectrum of the heavy singlets is
degenerate, namely $M_{1} = M_{2} =M_{3}=M_0 $, one can simply make
use of the tree-level matching condition
\begin{eqnarray}\label{eq:match}
\kappa^{}_{} |_{\mu}=  Y_\nu M^{-1}_{\rm R}  M_S (M^{T}_{\rm
R})^{-1} Y^T_\nu |_{\mu}\,
\end{eqnarray}
at the energy scale $\mu =M_0$. In the most general case with
nondegenerate heavy singlets, i.e., $M_1 < M_2 < M_3$, the
situation becomes more complicated and the heavy singlets have to be
sequentially decoupled from the theory~\cite{Antusch:2002rr}. For
instance, at energy scales between the $n$th and $(n-1)$th
thresholds, the heavy singlets are partially integrated out, leaving
only a $3 \times (n-1)$ submatrix in $Y_\nu$, which is
nonvanishing in the basis, where the heavy singlet mass matrix is
diagonal. The decoupling of the $n$th heavy singlet leads to the
appearance of an effective dimension-five operator similar to that
in Eq.~\eqref{eq:operator}, and the effective neutrino mass matrix
below $M_n$ is described by two parts
\begin{eqnarray}\label{eq:m-eff}
m^{(n)}_\nu = v^2 \left[ \kappa^{(n)}+Y_{\nu}^{(n)}\left(M_{\rm
R}^{(n)}\right)^{-1}M_{S}^{(n)}\left(M_{\rm
R}^{(n)T}\right)^{-1}Y_{\nu}^{(n)T}\right] \, ,
\end{eqnarray}
where $(n)$ labels the quantities relevant for the effective theory
between the $n$th and $(n-1)$th threshold. In the SM, the RGEs for
the two terms above have different coefficients for the gauge
coupling and Higgs self-coupling contributions, which can be traced
back to the decoupling of the right-handed neutrinos from the
theory. However, such a mismatch is absent in the MSSM due to the
supersymmetric structure of the MSSM Higgs and gauge
sectors.\footnote{We assume that the relevant threshold is above the
soft supersymmetry-breaking scale.} Therefore, this feature may result in
significant RG running effects only in the SM, in particular, when
the mass spectrum of the heavy neutrinos is quite hierarchical.

\section{Analytical RGEs for neutrino parameters}
\label{sec:approxRGEs}

In order to analytically investigate the RG evolution of leptonic
mixing angles, CP-violating phases, and neutrino masses, one can
translate the full RGEs for the neutrino mass matrix into a system
of differential equations for these parameters. The corresponding
formulas have been discussed below the seesaw
scale~\cite{Casas:1999tg,*Chankowski:1999xc,*Antusch:2003kp}, as
well as above the seesaw thresholds in the
type-I~\cite{Antusch:2005gp,*Mei:2005qp},
-II~\cite{Chao:2006ye,*Schmidt:2007nq}, and
-III~\cite{Chakrabortty:2008zh} seesaw models.

In general, there are two different strategies for solving the
resulting system. In the {\it top-down approach}, the initial
condition is specified at a certain high-energy scale, often motivated
by the flavor structure of a specific GUT-scale scenario.  This is an
advantage, since all the necessary ingredients are fixed at the high-energy scale and the running down to the electroweak scale is a mere
tedium. On the other hand, only some regions in the parameter space of
the full theory would presumably lead to good fits of the low-energy
data, which, together with an {\it ab initio} model dependency, makes
this method potentially quite inefficient.

In this study, we instead adopt the {\it bottom-up approach}, in which
the initial condition for the observables of our interest is fixed at
the low-energy scale, thus exploiting all the available experimental
information from the beginning. It is also clear from the mere
parameter counting that at the matching scale one can determine the
underlying theory parameters only up to an equivalence class defined
by the matching condition, and thus, the model-dependency problem is
somewhat delayed. Nevertheless, in order to set off from the
threshold, a representative should be chosen, which amounts to adding
extra assumptions on the flavor structure of the underlying theory at
the matching point. For a more detailed discussion on these issues,
see e.g.~Ref.~\cite{Davidson:2001zk} and references therein.

In particular, one can make use of some of the qualitative features
of the RGEs of the full theory. For instance, in the type-I seesaw
model, the structure of Eqs.~\eqref{eq:RGE1}-\eqref{eq:RGE4} and
\eqref{eq:RGEkappa} justifies the typical diagonality assumption
made on $Y_\nu $.  Indeed, if both $Y_e$ and $Y_\nu$ are diagonal at
a common energy scale, then they will retain the one-loop
diagonality at any energy scale, since there is no nondiagonal
element in the RGEs for $Y_e$ and $Y_\nu$. Thus, for the sake of
simplicity, we assume $Y_\nu = {\rm diag}
(y_{\nu1},y_{\nu2},y_{\nu3})$ in the basis in which $Y_{e}$ is also
diagonal,\footnote{Note that the simplifying assumption of a
simultaneous diagonality of $Y_{e}$, $Y_{\nu}$, and $M_{{\rm R}}$ is
specific for the inverse seesaw model and cannot be imposed in
e.g.~the type-I seesaw model.} which is reflected by the fact that
from now on we work with just three combinations of the leptonic
Yukawa couplings, namely
\begin{equation}\label{Yukawas}
(y^2_1,y^2_2,y^2_3) = C^e_\nu (y^2_e,y^2_\mu,y^2_\tau) + C_m
(y^2_{\nu 1},y^2_{\nu 2},y^2_{\nu 3}).
\end{equation}
We will briefly comment on the general case with a nondiagonal
Yukawa coupling matrix $Y_\nu$ later in Sec.~\ref{sect:genericY}.
The leptonic mixing matrix corresponds then to the unitary matrix
$U$ diagonalizing $\kappa$ as
\begin{eqnarray}\label{eq:U}
\kappa = U {\rm diag} (k_1,k_2,k_3) U^T \, ,
\end{eqnarray}
where $k_i$ ($i=1,2,3$) are the eigenvalues of $\kappa$. It is usually
given using the standard parametrization
\begin{widetext}
\begin{eqnarray}\label{eq:para}
U & = & P_\phi \left( \begin{matrix}c_{12} c_{13} & s_{12} c_{13} &
s_{13} e^{-{\rm i}\delta} \cr -s_{12} c_{23}-c_{12} s_{23} s_{13}
e^{{\rm i} \delta} & c_{12} c_{23}-s_{12} s_{23} s_{13} e^{{\rm i}
\delta} & s_{23} c_{13} \cr s_{12} s_{23}-c_{12} c_{23} s_{13}
e^{{\rm i} \delta} & -c_{12} s_{23}-s_{12} c_{23} s_{13} e^{{\rm i}
\delta} & c_{23} c_{13}\end{matrix} \right) \left(
\begin{matrix} e^{{\rm i}\rho} & & \cr  & e^{{\rm i}\sigma} & \cr & &
1 \end{matrix} \right) \ ,
\end{eqnarray}
\end{widetext}
with $c_{ij} \equiv \cos \theta_{ij}$ and $s_{ij} \equiv \sin
\theta_{ij}$ ($ij=12$, $13$, $23$). Here $P_\phi = {\rm diag} (e^{{\rm
    i} \phi_1},e^{{\rm i} \phi_2},e^{{\rm i} \phi_3})$ denotes three
unphysical phases, which are required for the diagonalization of an
arbitrary complex symmetric matrix.

For the RG running of leptonic mixing angles and CP-violating
phases, in order to simplify the results, we define the quantities
\begin{eqnarray}
\zeta_{ij}=\frac{m_i - m_j}{ m_i + m_j} \, .
\label{eq:zetas}
\end{eqnarray}
Inserting Eqs.~\eqref{eq:U} and \eqref{eq:zetas} into
Eq.~\eqref{eq:RGEkappa} and using Eq.~\eqref{Yukawas}, we arrive after
some tedious calculations at the RGEs for the leptonic mixing angles
and the CP-violating phases in the current scheme. The explicit, but
rather cumbersome results, are shown in App.~\ref{app:exactRGEs}.

In the case that the mass spectrum of the light neutrinos is nearly
degenerate, i.e., $m_1 \simeq m_2 \simeq m_3$, one can expect large
enhancement factors $\zeta^{-1}_{ij}$, which strongly boost the RG
running effects. In particular, for the leptonic mixing angle
$\theta_{12}$, the main RG running effect comes from the correction
proportional to $\zeta^{-1}_{21}$, and we obtain approximately
\begin{eqnarray}\label{eq:theta12approx}
\dot\theta_{12} & \simeq & \frac{1}{\zeta _{21}}c_{\rho -\sigma }
\left\{ s_{12}  c_{12} c_{\rho -\sigma } \left[c_{13}^2
y_1^2+\left(s_{23}^2 s_{13}^2 -c_{23}^2\right) y_2^2
-\left(s_{23}^2-c_{23}^2 s_{13}^2\right) y_3^2\right]  \right.
\nonumber \\ && \left. + s_{23}c_{23}s_{13} \left(c_{\delta +\rho
-\sigma }-2 s_{12}^2 c_{\delta } c_{\rho -\sigma } \right)
\left(y_3^2-y_2^2\right)\right\} \, .
\end{eqnarray}
If further neglecting the small leptonic mixing angle $\theta_{13}$,
we arrive at
\begin{eqnarray}\label{eq:theta12approx2}
\dot\theta_{12} & \simeq & \frac{1}{\zeta _{21}} s_{12}  c_{12}
c^2_{\rho -\sigma } \left( y_1^2-c_{23}^2 y_2^2-s_{23}^2
y_3^2\right)\, .
\end{eqnarray}
In addition, for the leptonic mixing angles $\theta_{23}$ and
$\theta_{13}$, we only keep the terms, which are not suppressed by
$s_{13}$, and obtain approximately
\begin{eqnarray}\label{eq:theta23approx}
\dot\theta_{23} & \simeq &  \frac{1}{\zeta _{31}}s_{23}c_{23}
\left(s_{12}^2c_{\rho }^2 +c_{12}^2 c_{\sigma }^2\right)
\left(y_2^2-y_3^2\right) \, ,  \\
\label{eq:theta13approx}\dot\theta_{13} & \simeq &\frac{1}{\zeta
_{31}}s_{12} c_{12}s_{23} c_{23} \left(c_{\sigma } c_{\delta +\sigma
}-c_{\rho } c_{\delta +\rho }\right)\left(y_2^2-y_3^2\right) \, ,
\end{eqnarray}
where the approximate relation $\zeta^{-1}_{31} \simeq
\zeta^{-1}_{32}$ has been used.

Now, we turn to the analytical RGEs for the CP-violating phases. In
the limit $\theta_{13} \to 0$, it is worthwhile to mention that the Dirac
CP-violating phase $\delta$ loses its meaning. However, it has been
pointed out that, with the RG running, both nontrivial
$\theta_{13}$ and $\delta$ can be generated
radiatively~\cite{Joshipura:2002kj,*Joshipura:2002gr,*Mei:2004rn,*Dighe:2008wn}.
Therefore, we keep terms proportional to either the inverse power of
$s_{13}$ or $\zeta^{-1}_{21}$, and obtain
\begin{eqnarray}\label{eq:phase-approx1}
\dot\delta & \simeq & \frac{1}{s_{13} \zeta _{31}}s_{12}c_{12}
s_{23}c_{23} \left(c_{\rho } s_{\delta +\rho }-c_{\sigma } s_{\delta
+\sigma }\right) \left(y_2^2-y_3^2\right) \nonumber \\ &&
+\frac{1}{2 \zeta _{21}}s_{2 \rho -2 \sigma } \left(y_1^2-c_{23}^2
y_2^2-s_{23}^2 y_3^2\right) \, ,  \\
\label{eq:phase-approx2}\dot\rho & \simeq & \frac{1}{2 \zeta
_{21}}c_{12}^2 s_{2 \rho -2 \sigma } \left(c_{23}^2 y_2^2+s_{23}^2
y_3^2-y_1^2\right) \, ,  \\ \label{eq:phase-approx3} \dot\sigma &
\simeq & \frac{1}{2 \zeta _{21}}s_{12}^2 s_{2 \rho -2 \sigma }
\left(c_{23}^2 y_2^2+s_{23}^2 y_3^2-y_1^2\right) \, .
\end{eqnarray}

Finally, we express the analytical RGEs for the masses of the light
neutrinos as
\begin{eqnarray}\label{eq:mass-approx1}
\dot m_1 & \simeq & 2 m_1 \left[c_{12}^2 y_1^2+\left(c_{23}^2
y_2^2+s_{23}^2 y_3^2\right) s_{12}^2\right]+2 m_1 \alpha_{\nu } \, ,
\\ \label{eq:mass-approx2}
\dot m_2 & \simeq & 2 m_2  \left[s_{12}^2 y_1^2+\left(c_{23}^2
y_2^2+s_{23}^2 y_3^2\right) c_{12}^2\right]+2 m_2 \alpha_{\nu } \, ,
\\ \label{eq:mass-approx3}\dot m_3 & \simeq & 2 m_3 \left( s_{23}^2
y_2^2+c_{23}^2
y_3^2\right)+2 m_3 \alpha_{\nu} \, .
\end{eqnarray}
One can observe that the $y_i$'s play the key role in the RG running
of $\theta_{12}$. In fact, in the low-energy scale type-I seesaw
model, $y_{\nu i} \ll y_\tau $ has to be satisfied in order to
effectively suppress the masses of the light neutrinos, and therefore,
no visible RG running effects can be achieved in the SM.  However, in the
inverse seesaw model, the $y_i$'s can be chosen to be of order unity,
since the masses of the light neutrinos are diminished by the Majorana
insertion $M_S$ instead of $Y_\nu$, and therefore, sizable RG running
effects can be naturally expected in the inverse seesaw model.

Concerning the bottom-up approach adopted in this study, let us add
one more technical remark at this point. Upon crossing the seesaw
threshold, the matching between the full and the effective theories
can be very easily performed in a basis, where the mass matrix of the
heavy singlets is diagonal. However, since $M_{\rm R} \gg M_S$ is well
satisfied in the current framework, we can effectively work out the
matching in a basis, where $M_{\rm R}$ is diagonal. The inaccuracy
induced by this approximation is limited to ${\cal O}(M_S/M_{\rm R})$,
which can be safely ignored in our numerical calculations.

\section{Numerical analysis, illustrations, and interpretations}
\label{sec:numeric}

We proceed to the numerical evolution of the RGEs in order to show
the representative RG running behavior of the neutrino parameters in
the inverse seesaw model. In our computations, we solve the one-loop
RGEs exactly instead of using the approximate formulas. For this
purpose, we adopt the values of the neutrino mass-squared
differences $\Delta m^2_{ij}=m^2_i -m^2_j$ and the leptonic mixing
angles from a global fit of current experimental data in
Ref.~\cite{Schwetz:2008er} that we assume to be given at the energy
scale $\mu=M_Z$. Note that, since there is no compelling evidence of
a nonzero $\theta_{13}$ so far, we set $\theta_{13}=0$ throughout
the numerical illustrations, unless otherwise stated. We also use
the values of quark and charged-lepton masses as well as the gauge
couplings given in Ref.~\cite{Xing:2007fb}. In the SM case, we
choose the Higgs mass to be $m_H =140~{\rm GeV}$, and we assume the
shape of the Higgs spectrum to be driven by $\tan\beta=10$ and
$\tan\beta=50$ as well as $m_A= M_{Z}$ in the MSSM, where the latter
assumption is motivated by simplicity. Currently, there is no direct
experimental information on the absolute values of the masses of the
light neutrinos. However, the recent measurement on the cosmic
microwave background finds that the sum of the masses of the light
neutrinos is less than $0.58~{\rm eV}$ at 95~\%
C.L.~\cite{Komatsu:2010fb}. One can estimate that this constraint
can be satisfied if $m_i < 0.15~{\rm eV}$ ($i = 1,2,3$). In
addition, the sign of $\Delta m^2_{32} \simeq \Delta m^2_{31}$
remains undetermined.  Therefore, we have four representative
choices for the mass spectrum of the light neutrinos: normal and
hierarchical neutrino mass spectrum (NH) with $m_1\ll m_2 \ll m_3$;
normal and nearly degenerate neutrino mass spectrum (ND) with $m_1
\lesssim m_2 \lesssim m_3$; inverted and hierarchical neutrino mass
spectrum (IH) with $m_3 \ll m_1 \ll m_2 $; and inverted and nearly
degenerate neutrino mass spectrum (ID) with $m_3 \lesssim m_1
\lesssim m_2 $. As we have shown in both the NH and IH cases, there
is no visible enhancement factor, since $\zeta^{-1}_{21}
=\zeta^{-1}_{31} =\zeta^{-1}_{32} \simeq 1$ in the NH case and
$\zeta^{-1}_{21} =-\zeta^{-1}_{31} =-\zeta^{-1}_{32} \simeq 1$ in
the IH case. Thus, we will mainly work in the ND and ID cases in
order to gain sizable enhancement factors. For example, in the ND
case with $m_1 = 0.1~{\rm eV}$, $\zeta^{-1}_{21} \sim 500$ and
$\zeta^{-1}_{31} \simeq \zeta^{-1}_{32} \sim 20$ hold to a good
precision. As for the masses of the heavy singlets, it has been
pointed out that, if the masses are around the TeV scale, one may
search for their signatures at the LHC via the trilepton channels,
i.e., $pp\to \ell \ell \ell \nu$, in which the SM background is
relatively small~\cite{delAguila:2007em}. We thereby choose these
masses located around the TeV scale in our numerical studies. In
practice, we first consider the simplest situation with all masses
of the heavy singlets being identical, and then come to the most
general case with a nondegenerate mass spectrum of the heavy
neutrinos.

However, for such a low seesaw scale, one has to consider further
phenomenological constraints on the parameter space of the model. In
particular, significant nonunitarity effects in the leptonic mixing
can emerge in the TeV scale inverse seesaw if there are ${\cal
O}(1)$ Yukawa couplings in the lepton
sector~\cite{Malinsky:2009gw,Malinsky:2009df}. This is traced back
to an effective dimension-six operator
\begin{eqnarray}\label{eq:d=6}
{\cal L}_{\nu}^{d=6} = c \left(\overline{\ell_{\rm L}} \tilde\phi\right)
{\rm i} \slc\partial \left({\tilde\phi}^\dagger \ell_{\rm L}\right)
\, ,
\end{eqnarray}
where $ c \equiv \left(Y_\nu M^{-1}_{\rm R}\right) \left(Y_\nu
M^{-1}_{\rm R}\right)^\dagger $ at leading order. At the electroweak
scale, a noncanonical kinetic term for light neutrinos is
generated, which, after canonical normalization, results in a
nonunitary relation between the flavor and mass eigenstates $
N=\left(1-FF^\dagger\right/2) U $, where $U$ is a unitary matrix and
$F=v Y_\nu M^{-1}_{\rm R}$. For $|F| \geq {\cal
  O}(0.1)$, non-negligible non-unitarity effects could be visible in
the near detector of a future neutrino factory, in particular in the
$\nu_\mu \to \nu_\tau$ channel, and there are further constraints
coming from the universality tests of weak interactions, rare
leptonic decays, the invisible $Z$ width, and neutrino oscillation
data~\cite{Antusch:2008tz}. Thus, in what follows, we restrict
ourselves to $|Y_\nu| \lesssim 0.3$, ensuring full compatibility
with the nonunitarity constraints.

\subsection{Single threshold}

\begin{figure}[t]
\begin{center}\vspace{0.cm}
\includegraphics[width=16cm]{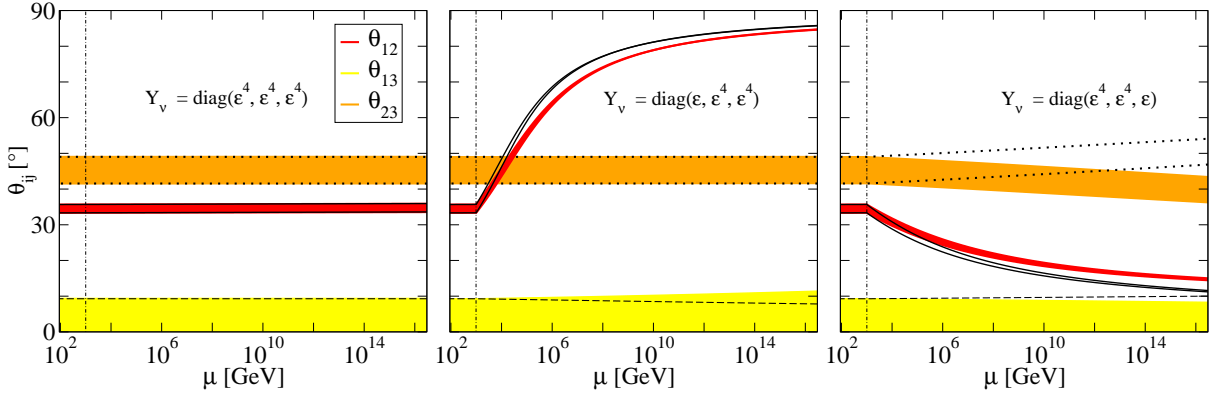}
\caption{\label{fig:fig1} The RG evolution behavior of the three
leptonic mixing angles as functions of the energy scale $\mu$ in the
SM. We use $M_0=1~{\rm TeV}$ and $\varepsilon=0.3$ as input
parameters at the seesaw scale. The colored regions correspond to
values of the leptonic mixing angles within their 1$\sigma$ confidence intervals in
the ND case, while the solid, dashed, and dotted curves denote the
same confidence intervals in the ID case.} \vspace{-0.cm}
\end{center}
\end{figure}

\begin{figure}[t]
\begin{center}\vspace{0.cm}
\includegraphics[width=16cm]{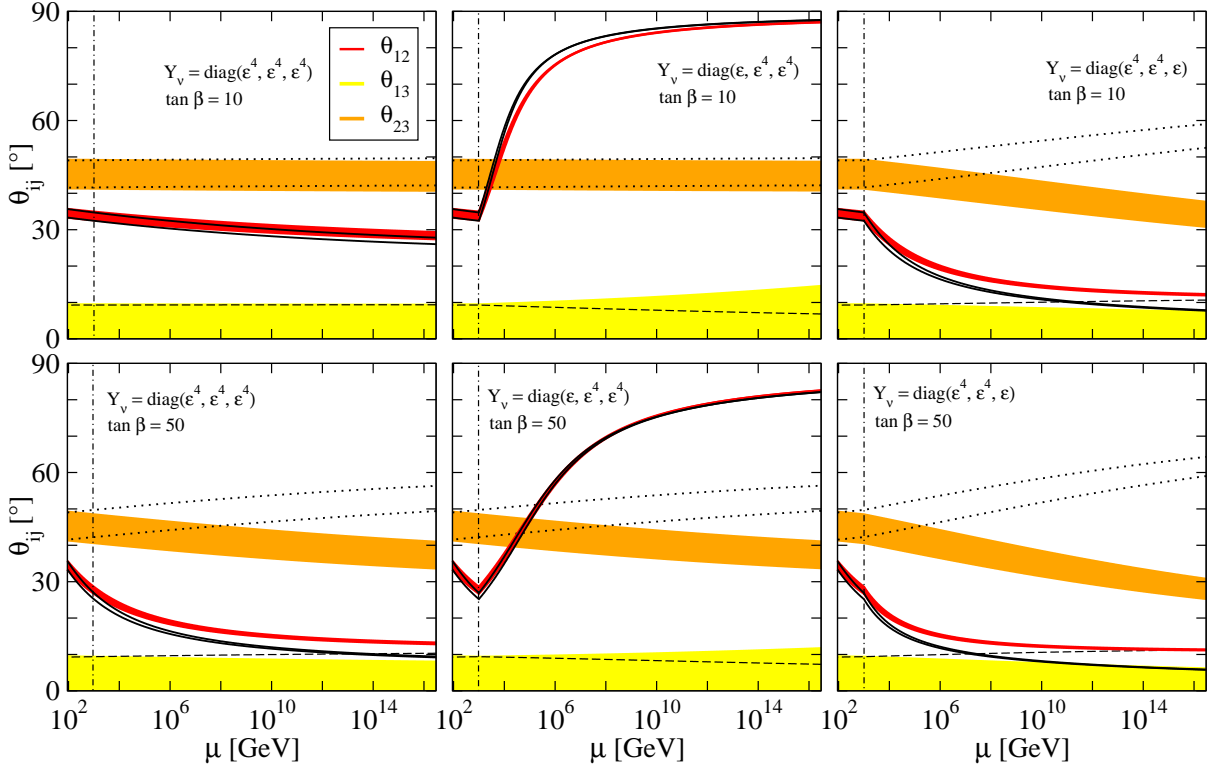}
\caption{\label{fig:fig2} The RG evolution behavior of the three
leptonic mixing angles as functions of the energy scale $\mu$ in the
MSSM with $\tan\beta=10$ (upper plots) and $\tan\beta=50$ (lower
plots), respectively. The input values of the masses of the heavy
singlets and the Yukawa couplings are the same as those in
Fig.~\ref{fig:fig1}.} \vspace{-0.cm}
\end{center}
\end{figure}

For the purpose of illustration, we consider the representative
example with $M_1 \simeq M_2 \simeq M_3 \simeq M_0 = 1 ~{\rm TeV}$,
called the {\it single threshold}. In Fig.~\ref{fig:fig1}, the RG
evolution behavior of the three leptonic mixing angles in the SM are
shown, in which the input values of $Y_\nu$ at $\mu = M_0$ are
labeled in each plot. Here, we do not consider the impact stemming
from the CP-violating phases, but we will comment on that later.
The left plot of Fig.~\ref{fig:fig1} shows that, in the case of small
Yukawa couplings, e.g. $|Y_\nu| \lesssim 0.01$, there is no visible
RG running effects on the leptonic mixing angles both below and
above the seesaw threshold. This can be observed from our analytical
formulas, since in the limit $y_{i} \to 0$, $\dot\theta_{ij} \simeq
0$ hold for all the three leptonic mixing angles. Note that the
meaning of $\varepsilon^4$ in the legends of
Figs.~\ref{fig:fig1}-\ref{fig:fig4} is that the corresponding
element is basically negligible compared with elements that are
indicated by $\varepsilon$. In the middle plot of
Fig.~\ref{fig:fig1}, one of the Yukawa couplings in $Y_\nu$ is
turned on, i.e., $y_{1} \simeq 0.3$, which leads to a significant
increase of $\theta_{12}$ at the GUT scale. On the other hand, if
$y_{1}$ is switched off, while $y_2$ or $y_{3}$ is turned on, as is
shown in the right plot of Fig.~\ref{fig:fig1}, $\theta_{12}$ will
decrease with increasing energy scale due to the opposite sign in
front of $y_1$ and $y_2$ or $y_3$ in Eq.~\eqref{eq:theta12approx2}.
From Fig.~\ref{fig:fig1}, we also find that the RG running of
$\theta_{12}$ is qualitatively insensitive to the hierarchy of the
masses of the light neutrinos. This is in agreement with our
analytical results, since $\theta_{12}$ mainly receives corrections
from $\zeta^{-1}_{21}$, which is positive for both possible
hierarchies. Furthermore, there exist quasifixed points at $\pi/2$
or 0 in the RG running of $\theta_{12}$. Such a feature can be
observed from Eq.~\eqref{eq:theta12approx2}, where $\dot\theta_{12}$
is proportional to both $s_{12}$ and $c_{12}$, which approaches zero
when $\theta_{12} \sim 0$ or $\theta_{12} \sim \pi/2$. Finally, due
to the lack of a sufficiently large enhancement factor,
$\theta_{23}$ and $\theta_{13}$ are relatively stable against
radiative corrections.  In principle, a shift of a few degrees for
$\theta_{23}$ can be achieved, see, for example, the right plot of
Fig.~\ref{fig:fig1}. In the limit $y_2 \simeq y_3$, their
contributions to the RG running of $\theta_{23}$ and $\theta_{13}$
are canceled, which can also be seen from
Eqs.~\eqref{eq:theta23approx} and \eqref{eq:theta13approx}.

In Fig.~\ref{fig:fig2}, we continue to illustrate the RG running of
the leptonic mixing angles in the MSSM for both small and large values
of $\tan\beta$.
Compared with the SM, the evolution of $\theta_{12}$ is analogous when
the RG running is dominated by $Y_\nu$. For instance, in the middle
and right columns of Fig.~\ref{fig:fig2}, $\theta_{12}$ approaches
gradually its quasi-fixed points in both the small and large
$\tan\beta$ cases. However, if $Y_\nu$ is suppressed, then the leading
contributions to the RG running of $\theta_{12}$ originate from the
charged-lepton Yukawa coupling $y_\tau$, in particular when
$\tan\beta$ is sizable (see, for example, the left column of
Fig.~\ref{fig:fig2}). This is similar to the RG evolution in the
effective theory, where the charged-lepton corrections are enhanced by
$\tan\beta$. Since $Y_e$ and $Y_\nu$ enter
Eq.~\eqref{eq:theta12approx2} with different signs, there is a
cancellation between them, namely, the $Y_\nu$ contributions to the RG
running are canceled somewhat by $Y_e$, and vice versa. In addition,
the RG running of $\theta_{23}$ and $\theta_{13}$ may be modified by
$y_\tau$, depending on the specific choice of the Yukawa couplings and
the value of $\tan\beta$. From the lower-right plot in
Fig.~\ref{fig:fig2}, one can observe that $\theta_{23}$ might acquire
sizable corrections if $\tan\beta$ is large. In order for
$\theta_{13}$ to have some visible RG running effects, arrangements of
the CP-violating phases have to be incorporated, otherwise
$\dot\theta_{13} \simeq 0$ according to
Eq.~\eqref{eq:theta23approx}. Therefore, the tiny RG running effects
in Fig.~\ref{fig:fig2} come from subleading order corrections.

The main difference between the RG running of the leptonic mixing
angles in the SM and the MSSM is that the $y_\tau$ contributions to
the running are negligible in the SM, while in the MSSM, they are
amplified by $\tan\beta$. For example, in the case of
$\tan\beta=50$, $y_\tau \simeq m_\tau /(v \cos\beta) \sim 0.3$,
indicating a substantial modification of the RG running. In the case
of $\tan\beta=10$, one can roughly estimate that $y_\tau \simeq
m_\tau /(v \cos\beta) \sim 0.1$, which leads to a relatively small
modification. This also reflects the fact that, in the MSSM, the RG
running of the leptonic mixing angles in the case of a small
$\tan\beta$ is weaker than that in the case of a large $\tan\beta$.
In addition, the discrepancies between the RG running in the ND and
ID cases originate from the sign changes in $\zeta^{-1}_{32}$ and
$\zeta^{-1}_{31}$ as shown in Eqs.~\eqref{eq:theta23approx} and
\eqref{eq:theta13approx}. For this reason, the RG running directions
of $\theta_{23}$ and $\theta_{13}$ are changed if the mass hierarchy
of the light neutrinos is reversed. Since the beta function of
$\theta_{12}$ is primarily dominated by $\zeta^{-1}_{12}$, the
running direction of $\theta_{12}$ is independent of the neutrino
mass hierarchy, although the subleading order corrections may
somewhat affect the RG running behavior.

\subsection{Multiple thresholds}
\label{sect:multiplethresholds}

\begin{figure}[t]
\begin{center}\vspace{0.cm}
\includegraphics[width=16cm]{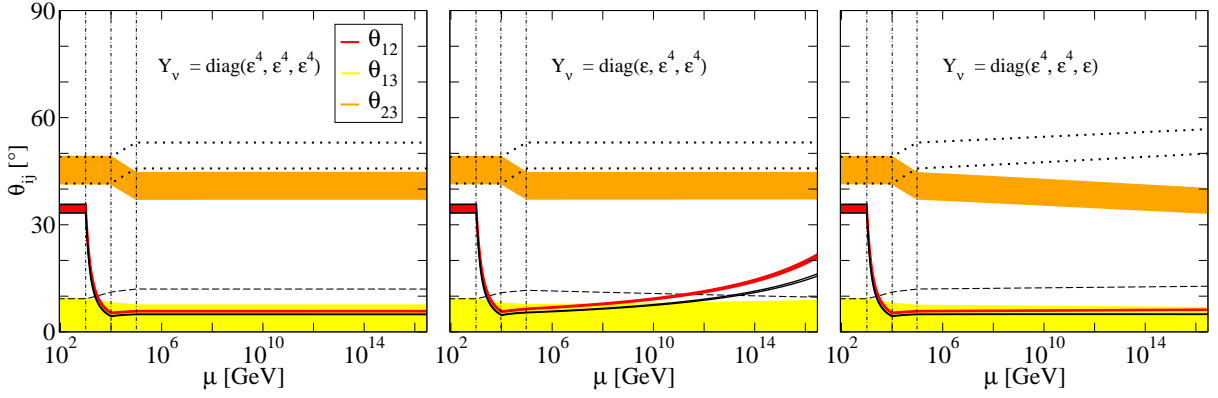}
\caption{\label{fig:fig3} The RG evolution behavior of the three
  leptonic mixing angles as functions of the energy scale $\mu$ in the
  SM with the masses of the heavy neutrinos being $1$, $10$, and
  $100~{\rm TeV}$, respectively. The input values of the Yukawa
  couplings are the same as those in
  Fig.~\ref{fig:fig1}.} \vspace{-0.cm}
\end{center}
\end{figure}

In the most general case with a nondegenerate mass spectrum of the
heavy singlets, i.e., $M_1 < M_2 < M_3$, the RG running effects
between the thresholds may modify the neutrino parameters remarkably
in the SM. This case is called the {\it multiple thresholds}. In
Fig.~\ref{fig:fig3}, we present the typical RG running behavior including
three thresholds at 1, 10, and 100~TeV in the SM.
Note that the chosen values of the three thresholds serve as an
order-of-magnitude example only. In crossing the thresholds from
$M_1$ to $M_3$, $\theta_{12}$ is dramatically suppressed no matter
the choice of $Y_\nu$. To identify the distinct threshold
corrections, we return to Eq.~\eqref{eq:m-eff}, and observe that the
neutrino mass matrix between the seesaw thresholds consists of two
parts $\kappa$ and $ Y_\nu M^{-1}_{\rm R} M_S (M^{T}_{\rm R})^{-1}
Y^T_\nu $. In the SM, the beta functions of these two parts have
different coefficients in the terms proportional to the gauge
couplings and the Higgs self-coupling
$\lambda$~\cite{Antusch:2005gp}. Therefore, keeping only the gauge
coupling and $\lambda$ corrections in the corresponding
beta functions, after a short-distance running from $M_{i}$ to
$M_{i-1}$, the two parts of the neutrino mass matrix are rescaled
to $a\kappa$ and $b Y_\nu M^{-1}_{\rm R} M_S (M^{T}_{\rm R})^{-1}
Y^T_\nu $, with $a\neq b$. Explicitly, we obtain the mass matrix of
the light neutrinos at $\mu=M_{i-1}$ as
\begin{eqnarray}\label{eq:mismatch}
m_\nu \Big|_{M_{i-1}} & \simeq & b v^2\left[\kappa +Y_\nu
M^{-1}_{\rm R} M_S (M^{T}_{\rm R})^{-1} Y^T_\nu \right]\Big|_{M_i} +
(a-b)v^2\kappa \Big|_{M_i} \nonumber \\ & = & b m_\nu \Big|_{M_{i}}
+ \Delta v^2\kappa \Big|_{M_i} \, ,
\end{eqnarray}
where
\begin{eqnarray}\label{eq:delta}
\Delta \simeq \frac{1}{16\pi^2}\left( \frac{3}{2}g^2_2 +
\frac{9}{10} g^2_1 + \lambda \right)
\ln\left(\frac{M_{i-1}}{M_i}\right)\, .
\end{eqnarray}
One may treat the second term in Eq.~\eqref{eq:mismatch} as a
perturbation, arising from the RG running between the different
thresholds. In contrast to the single threshold case, in which the RG
corrections to the leptonic mixing angles are due to the large Yukawa
couplings $Y_e$ and $Y_\nu$, the threshold corrections in this case
are related to the gauge couplings and $\lambda$. Their nontrivial
flavor structure is the reason for the observed RG running between the
thresholds. Furthermore, if the masses of the light neutrinos are
nearly degenerate, a short-distance running will result in sizable
corrections to the corresponding leptonic mixing angle due to the
enhancement factors $\zeta^{-1}_{ij}$.

The dramatic decrease of $\theta_{12}$ at around the first seesaw
threshold observed in Fig.~\ref{fig:fig3} is worth a further
comment. First, the discontinuity of the beta function at the seesaw
threshold is a direct consequence of the present renormalization
scheme, in which the heavy singlets are decoupled abruptly. If one
would instead use a mass-dependent scheme, the decoupling, and also
the observed RG running, would be smooth. Although the values of the
renormalized parameters in the vicinity of the thresholds are scheme
dependent, the total amount of running is less so. Second, the
negative slope of the curve between $M_{1}$ and $M_{2}$ is arguably
an artifact of the diagonality assumption imposed on the neutrino
Yukawa coupling matrix. This can clearly be observed from the
discussion in Sec.~\ref{sect:genericY} and seen in the corresponding
Fig.~\ref{fig:fig7} where an opposite behavior is observed in the
top-down approach with a nondiagonal Yukawa coupling matrix. Note
that a similar discussion applies also to the RG running behavior of
$\theta_{23}$ observed in Fig.~\ref{fig:fig3}.

Apart from the large variation at the seesaw thresholds, one may
expect a further increase of $\theta_{12}$ between $M_3$ and
the GUT scale (cf., the middle plot of Fig.~\ref{fig:fig3}),
although it is very unlikely for the two large leptonic mixing
angles $\theta_{23}$ and $\theta_{12}$ to unify at the GUT
scale,\footnote{Indicating a certain flavor symmetry in the lepton
sector at the given energy scale.} unless one increases the values
of the Yukawa couplings.  On the other hand, a smaller $\theta_{12}$
could be favored at the GUT scale, in particular a Cabibbo-like
angle, i.e., $\theta_C \sim 13^\circ$. In addition, $\theta_{23}$
acquires threshold corrections between the second and third
thresholds. Although they are milder compared with that of
$\theta_{12}$, the total amount of running between the thresholds
can be large if $M_3/M_2$ is much larger than what it is in our
example in Fig.~\ref{fig:fig3}.

\begin{figure}[t]
\begin{center}\vspace{0.cm}
\includegraphics[width=16cm]{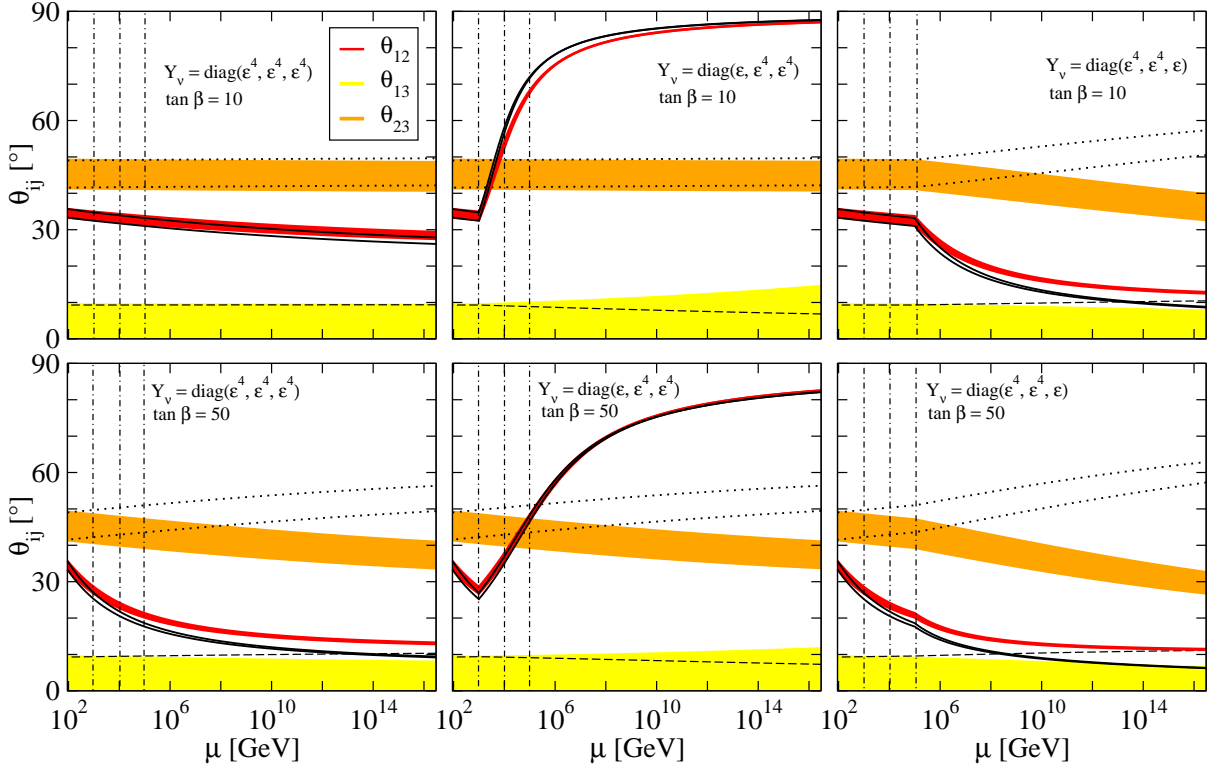}
\caption{\label{fig:fig4} The RG evolution behavior of the three
  leptonic mixing angles as functions of the energy scale $\mu$ in the
  MSSM with the masses of the heavy neutrinos being $1$, $10$, and
  $100~{\rm TeV}$, respectively. Similar to Fig.~\ref{fig:fig3}, we
  use $\tan\beta=10$ for the upper plots and $\tan\beta=50$ for the
  lower plots. The input values of the Yukawa couplings are the same
  as those in Fig.~\ref{fig:fig1}.} \vspace{-0.cm}
\end{center}
\end{figure}

In the MSSM, as we have shown in Sec.~\ref{sec:RGE}, there is no
mismatch between the RG running in the full and the effective theories.
Therefore, the running behavior of all three leptonic mixing angles
should be analogous to the ones in the single threshold case. The numerical
results are presented in Fig.~\ref{fig:fig4}, which are in agreement
with our expectations.

In conclusion, the RG evolutions of the leptonic mixing angles
between thresholds may lead to sizable corrections to the RG running
in the SM, owing to vertex renormalization. In general, one should
not simply use the matching condition at a common scale, unless the
mass spectrum of the heavy singlets is considerably degenerate or
the mass spectrum of the light neutrinos is very hierarchical (i.e.,
the NH or IH cases). Furthermore, we would like to comment on a
phenomenologically meaningful situation, in which two of the masses
of the heavy singlets are close to each other, while the other one
is not, e.g., $M_1 \simeq M_2 < M_3$ or $M_1 < M_2 \simeq
M_3$.\footnote{Such arrangements of the masses of the heavy singlets
may come from certain underlying flavor symmetries under which the
two heavy singlets transform as a doublet.} In such scenarios, the
threshold corrections to $\theta_{12}$ and $\theta_{13}$ are similar
to those in the general three threshold scenario. However, in the
latter case, $\theta_{23}$ is free of threshold effects, since it
only receives visible threshold corrections in the RG running
between $M_{2}$ and $M_3$, which can also be seen from
Fig.~\ref{fig:fig3}.

\subsection{Neutrino masses and CP-violating phases}

\begin{figure}[t]
\begin{center}\vspace{0.cm}
\includegraphics[width=6.5cm]{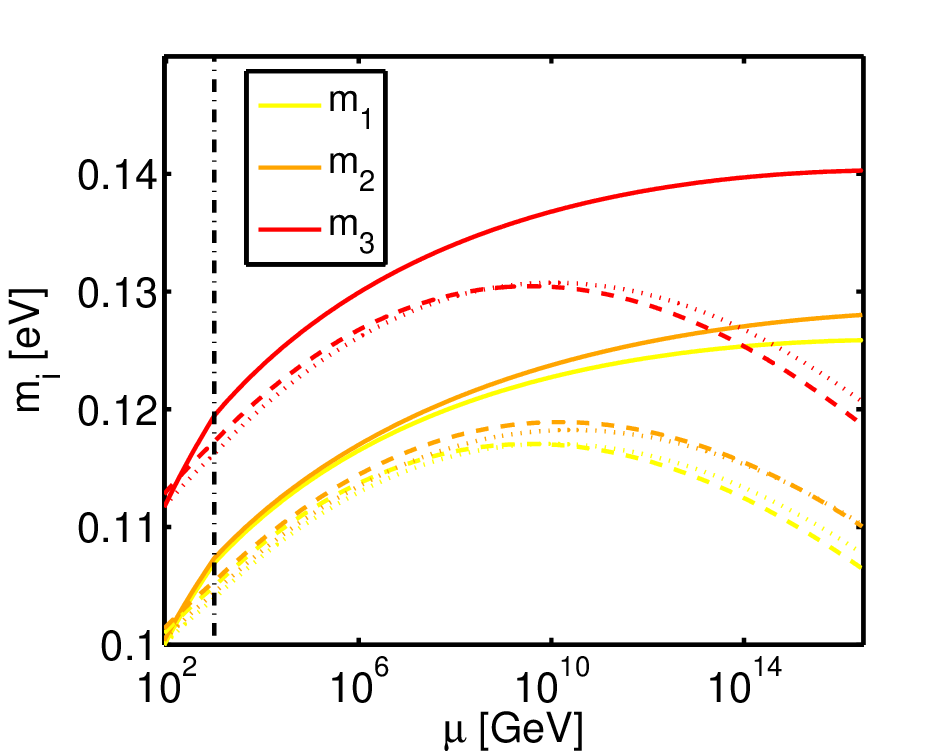}
\includegraphics[width=6.5cm]{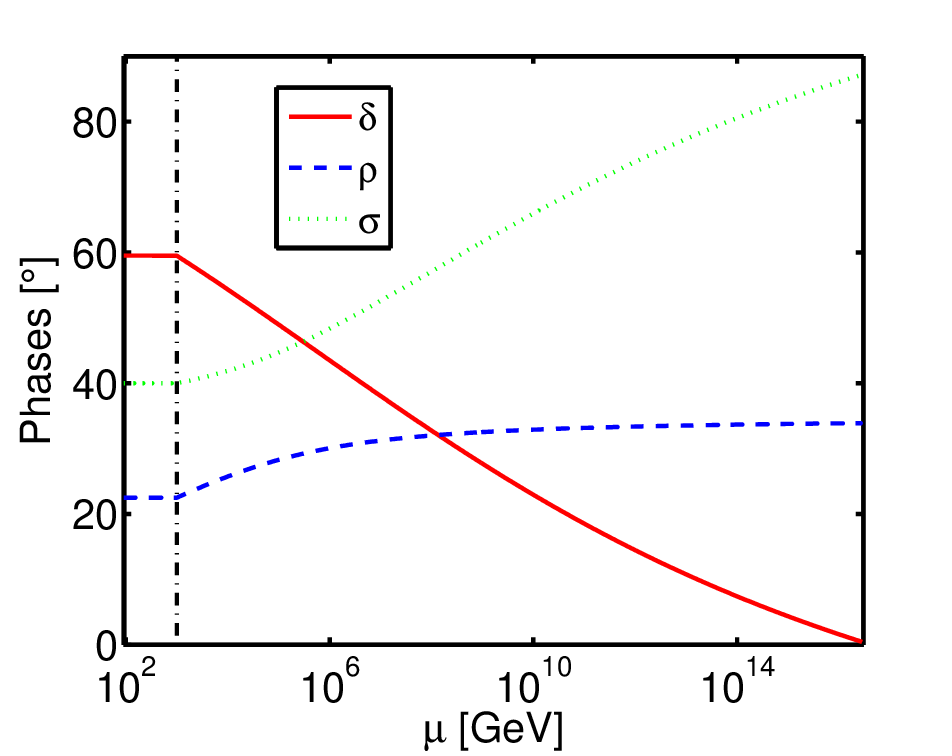}
\caption{\label{fig:fig5} The RG evolution behavior of the masses of
  the light neutrinos (left plot) and the CP-violating phases (right
  plot) as functions of the energy scale $\mu$ in the ND case. We use
  $Y_\nu ={\rm diag} (0.3,0.1,0.1)$ and $M_0 = 1~{\rm TeV}$ as an
  example. In the left plot, the solid, dashed, and dotted curves
  correspond to the cases of the SM, the MSSM with $\tan\beta =10$, and
  $\tan\beta=50$, respectively. In the right plot, we show the case of
  the SM only.} \vspace{-0.cm}
\end{center}
\end{figure}

For completeness, we perform the RG running of the neutrino masses
and the CP-violating phases, despite the fact that there is still a
lack of information on leptonic CP violation. As shown in
Fig.~\ref{fig:fig5} for the ND case, the RG running provides a
common rescaling of the masses of the light neutrinos, and there is
no sudden change along the running direction.
It can also be seen from
Eqs.~\eqref{eq:mass-approx1}-\eqref{eq:mass-approx3} that the RG
running of the masses of the light neutrinos is mainly governed by
the flavor blind part $\alpha_\nu$, while the $Y_\nu$ contributions
are relatively small. There exits a maximal value of the masses of
the light neutrinos in the MSSM at the energy scale $\mu\sim10^{10}
~{\rm GeV}$ due to the cancellation between gauge couplings and
Yukawa couplings in $\alpha_\nu$, while such a local maximum does
not appear in the SM. Note that similar RG running features of the
masses of the light neutrinos also exist in the ID case.

In the single threshold case, the RG running of the CP-violating
phases is also dominated by the corresponding Yukawa couplings.
According to Eqs.~\eqref{eq:phase-approx1}-\eqref{eq:phase-approx3},
there are enhancement factors for both the Dirac and Majorana
phases. Interestingly, even if one has a vanishing Dirac phase
$\delta$ at a certain high-energy scale, a nonvanishing $\delta$
can be generated radiatively via the Majorana phases. This can be
observed from the right plot of Fig.~\ref{fig:fig5}, where
$\delta=0$ is obtained at the GUT scale. As we have pointed out,
there is a subtlety at $\theta_{13}=0$, since the definition of
$\delta$ loses its meaning. Hence, we choose $s_{13}=0.01$ in our
realistic calculations. In addition, the Majorana phases $\rho$ and
$\sigma$ run in the same direction, since the coefficients of their
beta functions possess identical signs at leading order. A complete
analysis of the RG running of the CP-violating phases could be
interesting and useful for model building. However, such a study
lies beyond the scope of this work.

\subsection{Nondiagonal Yukawa couplings}
\label{sect:genericY}
\begin{figure}[t]
\begin{center}\vspace{0.cm}
\includegraphics[width=6.5cm]{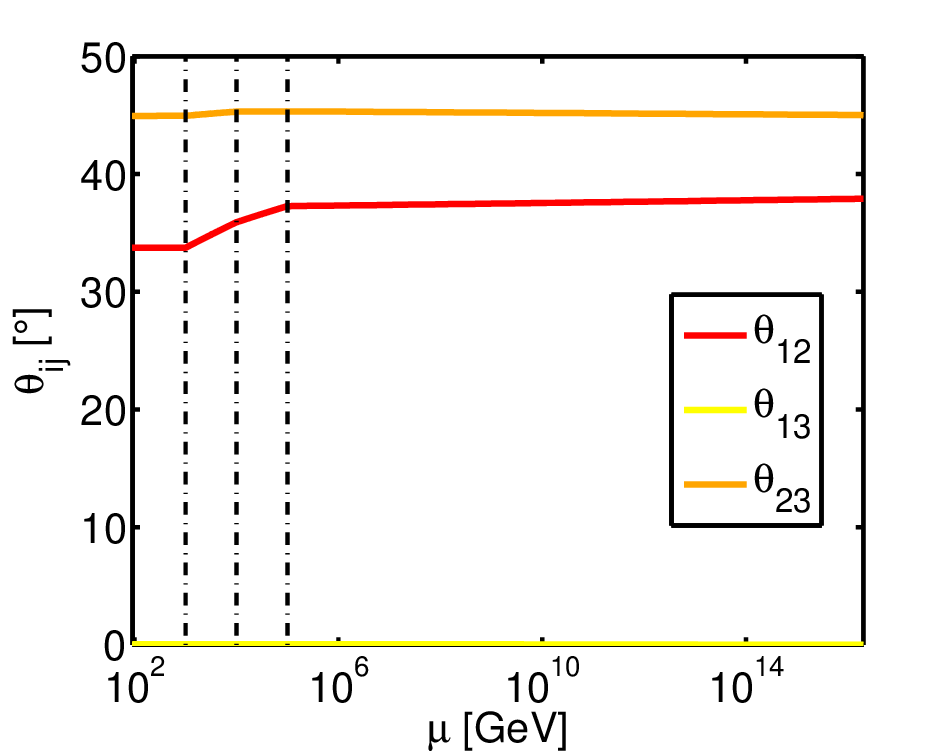}
\caption{\label{fig:fig6} {An example of the RG evolution of the
    leptonic mixing angles with a nondiagonal $Y_\nu$ in the SM. We
    focus on the NH case and use $m_1=0.01 ~ {\rm eV}$ as a sample
    value.}}
\vspace{-0.cm}
\end{center}
\end{figure}

Until now, our numerical analysis has been based on the assumption
that both $Y_e$ and $Y_\nu$ are diagonal at the GUT scale. For the
general case with arbitrary Yukawa couplings, it is very difficult to
obtain analytical RGEs for the neutrino parameters due to the
additional nonzero elements in $Y_\nu$. Thus, in this case, a
top-down approach is highly favorable, despite the technical challenge
of accommodating the low-energy data.

Not going into details, we focus on adding further support to the
comments made in Sec.~\ref{sect:multiplethresholds} on the
qualitative features of the behavior of $\theta_{12}$ and
$\theta_{23}$ among the seesaw thresholds in the SM as depicted in
Fig.~\ref{fig:fig3}.  In Fig.~\ref{fig:fig6}, we show a numerical
example corresponding to the NH case with an off-diagonal neutrino
Yukawa coupling matrix of the form
\begin{eqnarray}\label{eq:YnuE}
Y_\nu = \left(\begin{matrix}\varepsilon^4 & 0 & \varepsilon \cr 0 &
\varepsilon^4 & 0 \cr \varepsilon & 0 & \varepsilon^4
\end{matrix}\right) \, .
\end{eqnarray}
Unlike in the diagonal case of Fig.~\ref{fig:fig3}, one can find
directly that $\theta_{12}$ increases across the seesaw thresholds
and it can change significantly even between $M_{2}$ and $M_{3}$.
Similarly, in Fig.~\ref{fig:fig6}, $\theta_{23}$ runs in the
opposite direction compared with that in Fig.~\ref{fig:fig3} and
most of the running is experienced between $M_{1}$ and $M_{2}$. Note
also that the magnitude of the effects in Fig.~\ref{fig:fig6} is
smaller than in Fig.~\ref{fig:fig3}, since a smaller value of
$m_{1}$ is adopted for the sake of a better convergence of the top-down
approach. Therefore, in Fig.~\ref{fig:fig3}, the sharp decline of
$\theta_{12}$ is a consequence of the assumptions made on $Y_\nu$
rather than a generic feature.

\subsection{Neutrino mixing patterns and flavor symmetries}

\begin{figure}[t]
\begin{center}\vspace{0.cm}
\includegraphics[width=6.5cm]{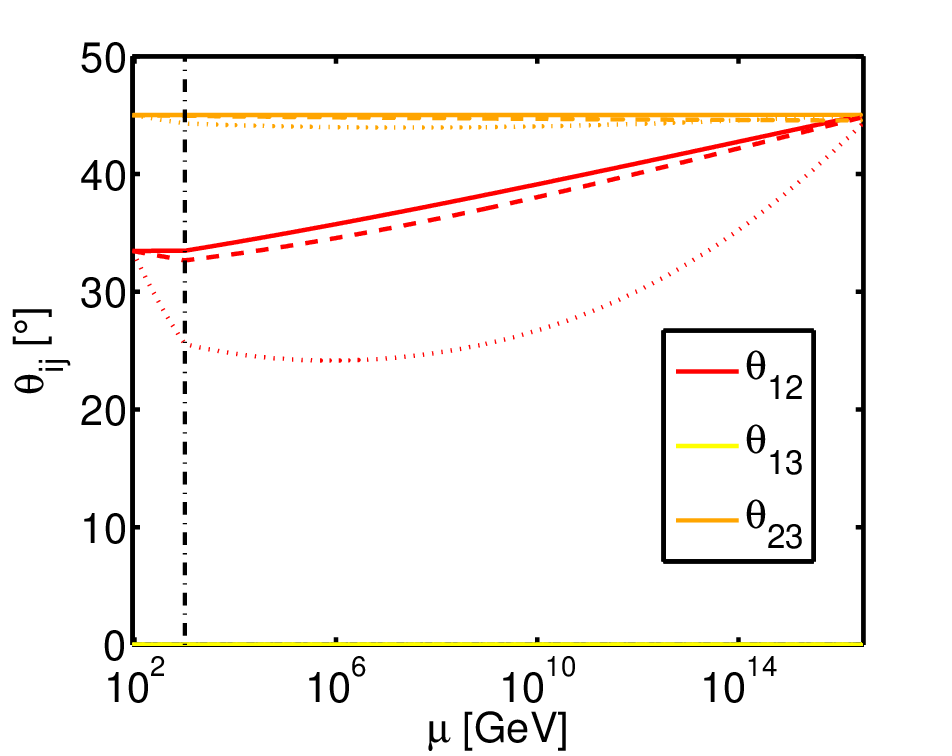}
\includegraphics[width=6.5cm]{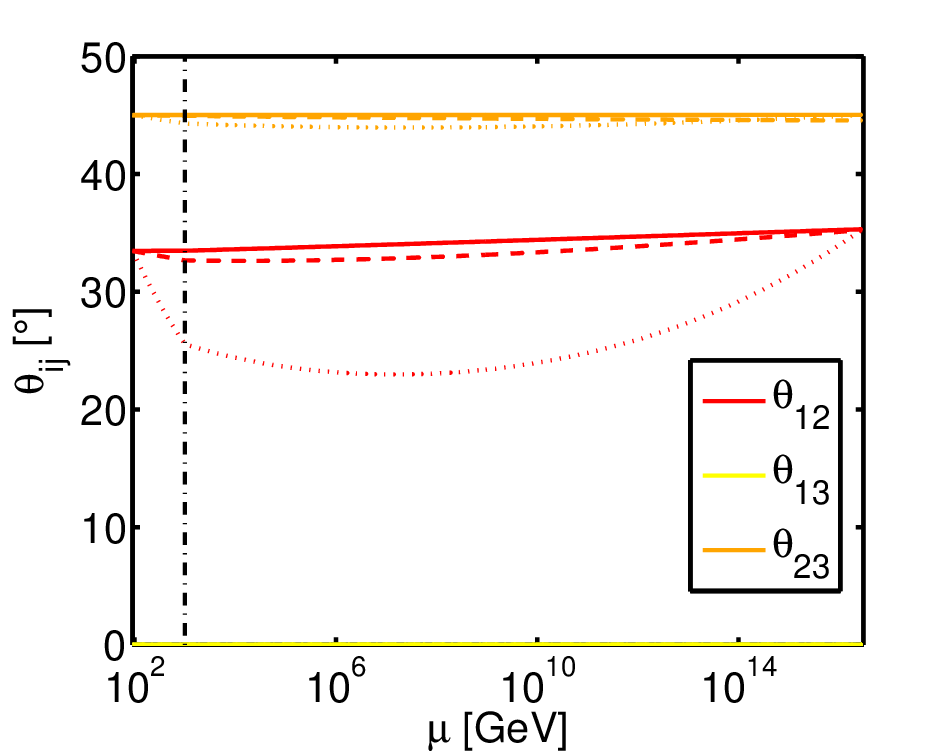}
\caption{\label{fig:fig7} Realizations of the bimaximal (left plot)
  and the tri-bimaximal (right plot) neutrino mixing patterns at the
  GUT scale. Here we only consider the ND case. In addition, for
  simplicity, we use $M_0 = 1~{\rm TeV}$ for all the heavy singlets
  and take all the CP-violating phases to be zero. The solid, dashed,
  and dotted curves represent the SM, the MSSM with $\tan\beta=10$ and
  $\tan\beta=50$, respectively.} \vspace{-0.cm}
\end{center}
\end{figure}

While understanding the flavor puzzle is a fundamental task in
particle physics, there are various flavor symmetric models
constructed at high-energy scales, where the flavor symmetry is
restored. In principle, any predictions from certain flavor
symmetries encounter radiative corrections, and therefore, it is
essential to take into account the RG running effects in determining
physical parameters at an observable energy scale. In the effective
theory, it is well known that the RG running effects are negligibly
small in the SM. In the MSSM with nearly degenerate mass spectrum of
the light neutrinos, $\theta_{12}$ decreases with increasing energy
scale. In view of these features in the effective theory, there is
no way to either unify $\theta_{12}$ and $\theta_{23}$ as suggested
in the bimaximal mixing pattern ($s_{12} =s_{23} =1/\sqrt{2}$ and
$s_{13}=0$ in the standard parametrization), or arrange them to the
tri-bimaximal mixing pattern ($s_{12} = 1/\sqrt{3}$,
$s_{23}=1/\sqrt{2}$, and $s_{13}=0$ in the standard parametrization)
at high-energy scales. Nevertheless, once the RG running effects
above the seesaw scale are included, it makes sense to realize
certain interesting mixing patterns at a unification scale. In
Fig.~\ref{fig:fig7}, we illustrate the possibility of realizing the
bimaximal and tri-bimaximal leptonic mixing patterns at the GUT
scale via fine-tuning of the Yukawa couplings in the RGEs.
For simplicity, we only show the single seesaw threshold case. As
examples, in the left plot of Fig.~\ref{fig:fig7}, we use
$$
(y_{\nu1},y_{\nu2},y_{\nu3}) = (0.083,0.010,0.010), ~ (0.085,0.040,0.010), ~ (0.266,0.260,0.010)
$$
in the SM, in the MSSM with $\tan\beta =10$, and in the MSSM with
$\tan\beta=50$, respectively, whereas in the right plot of
Fig.~\ref{fig:fig7}, we use
$$
(y_{\nu1},y_{\nu2},y_{\nu3}) =
(0.034,0.010,0.010), ~ (0.066,0.040,0.010),~ (0.261,0.260,0.010)
$$
in the SM, in the MSSM with $\tan\beta =10$, and in the MSSM with
$\tan\beta=50$, respectively. In all cases, flavor symmetric mixing
patterns are obtained according to the RG running. Thus various
mixing patterns can be naturally achieved by adjusting the Yukawa
couplings in the inverse seesaw model.

\section{Comparison with other seesaw models}
\label{sec:comparison}

As discussed in Sec.~\ref{sec:introduction}, there is no difference
between the RG evolutions in low-scale effective theories. If we stick
to the conventional type-I, -II, -III, and inverse seesaw models, at
energy scales above the seesaw threshold, the RG running of the
neutrino mass matrix can be uniformly described by
\begin{eqnarray}\label{eq:RGEall}
\dot m_\nu = N_\nu m_\nu + m_\nu N^T_\nu + \alpha_\nu m_\nu \, ,
\end{eqnarray}
where
\begin{eqnarray}\label{eq:Nall}
N_\nu = C_e Y_e Y^\dagger_e + C_A Y_A Y^\dagger_A \, ,
\end{eqnarray}
with $Y_A$ being the Yukawa couplings between heavy seesaw particles
and lepton doublets. The coefficients $\alpha_\nu$, $C_e$, and $C_A$
depend on the specific model. Apparently, sizable Yukawa couplings in
Eq.~\eqref{eq:Nall} could give birth to visible RG running effects.
In addition, threshold effects may induce significant corrections.

Let us now investigate these two points in detail. In the conventional
type-I seesaw model, larger Yukawa couplings mean higher energy scale
of the right-handed neutrinos, and thus, prominent RG running effects
could emerge at some super-high-energy scale far above the scope of
current experiments. If we lower the scale of the right-handed
neutrinos by reducing the corresponding Yukawa couplings, threshold
corrections may play an important role in the RG running. However,
such a theory is still lacking testability, since the interactions
between right-handed neutrinos and SM particles are suppressed by the
Yukawa couplings, unless we use severe fine-tuning or specific
assumptions on the model
parameters~\cite{Pilaftsis:1991ug,*Kersten:2007vk,*Antusch:2009gn,*Zhang:2009ac}.
In the type-II seesaw model, one may have sizable Yukawa couplings
without facing the problem of lacking observability. However, as shown
in Ref.~\cite{Chao:2006ye}, there is no enhancement factor in the
type-II seesaw framework, and one can hardly have visible RG running
effects. For the simplest type-II seesaw model, there is only one
triplet Higgs, which on the other hand prohibits the possibility of
threshold corrections. The situation in the type-III seesaw model is
similar to that in the type-I seesaw model, apart from the fact that,
due to the charged components, there could be visible collider
signatures no matter the magnitude of the Yukawa couplings.

The prominent feature of the RGEs in the inverse seesaw model is
that significant RG running of the neutrino parameters could occur
at lower energy scales without losing the testability at the current
experiments. The RG running does not spoil the stability of the
masses of the light neutrinos, but may introduce distinctive
corrections to the lepton flavor structure. Indeed, this is related
to the characteristic of the inverse seesaw model, namely, lepton
number violation is well separated from lepton flavor violation.

Note that, in all cases with visible RG running effects, a nearly
degenerate mass spectrum of the light neutrinos is required.
Otherwise, there is no efficient enhancement factor boosting the
running.  Since the beta functions in
Eqs.~\eqref{eq:mass-approx1}-\eqref{eq:mass-approx3} are
proportional to the masses of the light neutrinos explicitly, a
nonvanishing mass cannot be generated via the RG running if it is
zero at a certain energy scale.

\section{Summary and conclusions}
\label{sec:summary}

In this work, we have performed both analytical and numerical
analyses of the RG running of the neutrino parameters in the inverse
seesaw model. We have shown that, due to the sizable Yukawa
couplings between light neutrinos and heavy gauge singlets in the
inverse seesaw model, substantial RG running effects can be
naturally obtained even at low-energy scales. Such a running
distinguishes the inverse seesaw model from other simple seesaw
models, and may be experimentally tested in near-future experiments.
Concretely, we have derived very compact analytical RGEs for the
neutrino parameters above the seesaw scale. Furthermore, a detailed
numerical study of the RG running effects on the leptonic mixing
angles has been carried out in both the SM and the MSSM. In general,
we have found that there may be significant RG running effects on
$\theta_{12}$ and $\theta_{23}$, and in particular on $\theta_{12}$,
if the mass spectrum of the light neutrinos is nearly degenerate.
Furthermore, the running between the seesaw thresholds corresponding
to a certain hierarchy in the masses of the heavy neutrinos can be
strong. The RG running effects of a nondiagonal Yukawa coupling
matrix have also been briefly discussed. We have demonstrated that
some phenomenologically and theoretically interesting leptonic
mixing patterns, the bimaximal and tri-bimaximal patterns, can be
achieved at a high-energy scale once the RG running effects are
taken into account. In addition, the RG evolution of neutrino masses
and CP-violating phases has been studied qualitatively. We have
found that the Majorana phases run in the same direction and that a
nonzero Dirac phase at low-energy scales can be generated from a
vanishing one at some high-energy scale.

\begin{acknowledgments}
This work was supported by the Royal Swedish Academy of Sciences (KVA)
[T.O.], the G{\"o}ran Gustafsson Foundation [H.Z.], and the Swedish
Research Council (Vetenskapsr{\aa}det), Contract No.~621-2008-4210
[T.O.].
\end{acknowledgments}

\newpage

\appendix

\section{Complete RGEs for neutrino parameters}
\label{app:exactRGEs}

The complete one-loop RG running of the masses of the light
neutrinos are given by
\begin{eqnarray}\label{eq:m}
\dot m_1 & = & 2 m_1 \left\{c_{12}^2\left[s_{13}^2\left(s_{23}^2
y_2^2+c_{23}^2 y_3^2\right) +c_{13}^2 y_1^2\right] +2 s_{12} c_{12}
s_{23} c_{23}s_{13}c_{\delta}\left(y_2^2-y_3^2\right) \right.
\nonumber \\ && \left. +s_{12}^2 \left(c_{23}^2 y_2^2+s_{23}^2
y_3^2\right)+\alpha _{\nu }\right\} \, ,
\\ \dot m_2 & = & 2 m_2
\left[c_{12}^2\left(c_{23}^2 y_2^2+s_{23}^2 y_3^2\right)+2 s_{12}
c_{12} s_{23}c_{23} s_{13}c_{\delta
} \left(y_3^2-y_2^2\right) +s_{12}^2 c_{13}^2 y_1^2\right. \nonumber \\
&& \left.+s_{12}^2 s_{13}^2 \left(s_{23}^2 y_2^2+c_{23}^2
y_3^2\right)+\alpha _{\nu }\right] \, , \\ \dot m_3 & = & 2
m_3\left\{ \left[c_{13}^2\left(s_{23}^2 y_2^2+c_{23}^2 y_3^2\right)
+s_{13}^2 y_1^2\right]+ \alpha _{\nu }\right\} \, ,
\end{eqnarray}
where we have defined $\dot\theta_{ij} \equiv 16\pi^2 \mu \frac{{\rm
d}\theta_{ij}}{{\rm d}\mu}$, $c_\delta \equiv \cos\delta$,
$s_{\delta-\rho} \equiv \sin(\delta-\rho)$, and so on. Similarly,
the full analytical results for the leptonic mixing angles and the Dirac CP-violating phase read
\begin{eqnarray}\label{eq:theta12}
\dot\theta_{12} & = & \zeta _{21}s_{\rho -\sigma }
\left\{s_{23}c_{23} s_{13}\left(2 c_{\delta } s_{12}^2 s_{\rho
-\sigma }-s_{\delta +\rho -\sigma }\right) \left(y_2^2-y_3^2\right)
\right. \nonumber \\ && \left. ~~~ + s_{12}c_{12}s_{\rho -\sigma }
\left[c_{13}^2 y_1^2+\left(s_{23}^2s_{13}^2 -c_{23}^2\right)
y_2^2-\left(s_{23}^2-c_{23}^2
s_{13}^2\right) y_3^2\right]\right\}\nonumber \\
&+ & \zeta _{31}s_{12} s_{13} s_{\delta +\rho } \left\{s_{12} s_{23}
c_{23} s_{\rho } \left(y_3^2-y_2^2\right)+c_{12} s_{13} s_{\delta
+\rho } \left[\left(y_3^2-y_2^2\right)
s_{23}^2+y_1^2-y_3^2\right]\right\} \nonumber \\
&+& \zeta _{32}s_{13} s_{\delta +\sigma }
\left\{s_{23}c_{23}s_{\sigma } \left(y_3^2-y_2^2\right)
c_{12}^2-s_{12} c_{12}s_{13} s_{\delta +\sigma
} \left[\left(y_3^2-y_2^2\right) s_{23}^2+y_1^2-y_3^2\right]\right\} \nonumber \\
&+&\frac{1}{\zeta _{21}}c_{\rho -\sigma }
\left\{s_{23}c_{23}s_{13}\left(c_{\delta +\rho -\sigma }-2 c_{\delta
} c_{\rho -\sigma } s_{12}^2\right)\left(y_3^2-y_2^2\right) \right.
\nonumber \\ && \left. ~~~ +s_{12}c_{12} c_{\rho -\sigma
}\left[c_{13}^2 y_1^2+\left(s_{23}^2s_{13}^2 -c_{23}^2\right)
y_2^2-\left(s_{23}^2-c_{23}^2
s_{13}^2\right) y_3^2\right]\right\} \nonumber \\
&+&\frac{1}{\zeta _{31}}c_{\delta +\rho } s_{12} s_{13}
\left\{s_{12} s_{23}c_{23} c_{\rho
}\left(y_3^2-y_2^2\right)+c_{12}s_{13}c_{\delta
+\rho } \left[\left(y_3^2-y_2^2\right) s_{23}^2+y_1^2-y_3^2\right]\right\} \nonumber \\
&+&\frac{1}{\zeta _{32}}c_{\delta +\sigma } s_{13}
\left\{c_{12}^2s_{23}c_{23} c_{\sigma }\left(y_3^2-y_2^2\right) -
s_{12} c_{12}s_{13} c_{\delta +\sigma
}\left[\left(y_3^2-y_2^2\right) s_{23}^2+y_1^2-y_3^2\right]\right\} \, ,
\end{eqnarray}
\begin{eqnarray}\label{eq:theta23}
\dot\theta_{23} & = &
\zeta _{31}s_{12} s_{\rho } \left\{c_{23} s_{12} s_{23} s_{\rho }
\left(y_2^2-y_3^2\right)-c_{12} s_{13} s_{\delta +\rho }
\left[\left(y_3^2-y_2^2\right)
s_{23}^2+y_1^2-y_3^2\right]\right\} \nonumber \\
&+&\zeta _{32}s_{\sigma } \left\{c_{12}^2s_{23}c_{23}{ }s_{\sigma }
\left(y_2^2-y_3^2\right) +s_{12}c_{12} s_{13} s_{\delta +\sigma }
\left[\left(y_3^2-y_2^2\right)
s_{23}^2+y_1^2-y_3^2\right] \right\} \nonumber \\
&+&\frac{1}{\zeta _{31}}c_{\rho } s_{12} \left\{s_{12} s_{23} c_{23}
c_{\rho }\left(y_2^2-y_3^2\right)-c_{12} s_{13}c_{\delta +\rho }
\left[\left(y_3^2-y_2^2\right)
s_{23}^2+y_1^2-y_3^2\right]\right\} \nonumber \\
&+&\frac{1}{\zeta _{32}}c_{\sigma } \left\{c_{12}^2 s_{23} c_{23}
c_{\sigma }\left(y_2^2-y_3^2\right) + s_{12}c_{12} s_{13} c_{\delta
+\sigma }\left[\left(y_3^2-y_2^2\right) s_{23}^2+y_1^2-y_3^2\right]
\right\} \, ,
\end{eqnarray}
\begin{eqnarray}\label{eq:theta13}
\dot\theta_{13} & = & \zeta _{31}c_{13} s_{\delta
+\rho } \left\{c_{12}^2s_{13} s_{\delta +\rho }
\left[\left(y_3^2-y_2^2\right) s_{23}^2+y_1^2-y_3^2\right] + s_{12}
c_{12}s_{23}c_{23} s_{\rho } \left(y_3^2-y_2^2\right) \right\} \nonumber \\
&+&\zeta _{32}s_{12}c_{13}\text{  }s_{\delta +\sigma } \left\{c_{12}
s_{23} c_{23}s_{\sigma } \left(y_2^2-y_3^2\right)+s_{12} s_{13}
s_{\delta
+\sigma } \left[\left(y_3^2-y_2^2\right) s_{23}^2+y_1^2-y_3^2\right]\right\} \nonumber  \\
&+&\frac{1}{\zeta _{31}}c_{13} c_{\delta +\rho }
\left\{c_{12}^2s_{13} c_{\delta +\rho }
\left[\left(y_3^2-y_2^2\right) s_{23}^2+y_1^2-y_3^2\right]
+ s_{12}c_{12} s_{23}c_{23} c_{\rho } \left(y_3^2-y_2^2\right) \right\} \nonumber \\
&+&\frac{1}{\zeta _{32}}s_{12}c_{13} c_{\delta +\sigma
}\left\{c_{12} s_{23}c_{23} c_{\sigma }\left(y_2^2-y_3^2\right)+
s_{12} s_{13}c_{\delta +\sigma} \left[\left(y_3^2-y_2^2\right)
s_{23}^2+y_1^2-y_3^2\right]\right\} \, ,
\end{eqnarray}
\begin{eqnarray}\label{eq:RGEdelta}
\dot\delta & = &  \frac{\zeta _{21}}{2 c_{12} c_{23} s_{12}}\left\{2
c_{\rho -\sigma } s_{13} s_{23} \left(s_{\delta +\rho -\sigma }-2
c_{\delta } s_{12}^2 s_{\rho -\sigma }\right)
\left(y_2^2-y_3^2\right) c_{23}^2\right. \nonumber \\
&& + \left.c_{12} s_{12} s_{2 \rho -2 \sigma } \left[-c_{13}^2
y_1^2-\left(s_{13}^2 s_{23}^2-c_{23}^2\right)
y_2^2+\left(s_{23}^2-c_{12}^2 s_{23}^2\right)
y_3^2\right] c_{23}\right\} \nonumber \\
&+&\frac{\zeta _{31}}{2 c_{12} c_{23} s_{13} s_{23}}
 \left\{c_{12} c_{23} s_{13} s_{23} \left[\left(\left(s_{23}^2
\left(s_{2 \delta +2 \rho } s_{13}^2+2 s_{2 \rho } +s_{2 \delta +2
\rho }\right)-s_{2 \rho }\right) y_2^2 \right.\right.\right. \nonumber \\
&& \left. + \left(s_{2 \delta +2 \rho } s_{13}^2+ 2 c_{\delta }
s_{\delta +2 \rho }-s_{23}^2 \left(s_{2 \delta +2 \rho } s_{13}^2+2
s_{2 \rho }+s_{2 \delta +2 \rho }\right)\right) y_3^2\right)
s_{12}^2 \nonumber \\ && - \left.\left(s_{12}^2
\left(s_{13}^2+1\right)-1\right) s_{2 \delta +2 \rho } y_1^2-s_{2
\delta +2 \rho } \left(\left(y_2^2-y_3^2\right)
s_{23}^2+y_3^2\right)\right]  \nonumber \\ && + s_{12}
\left[s_{13}^2 \left(\left(4 c_{\rho } s_{\delta +\rho }-s_{12}^2
\left(s_{\delta }+3 s_{\delta +2 \rho }\right)\right)
\left(y_2^2-y_3^2\right) s_{23}^4 \right.\right. \nonumber \\ &&  +
\left(\left(2 s_{12}^2 s_{\delta +2 \rho }-2 c_{\rho } s_{\delta
+\rho }\right) y_2^2+2 \left(3 c_{\rho } s_{\delta +\rho }-s_{12}^2
\left(s_{\delta }+2
s_{\delta +2 \rho }\right)\right) y_3^2\right) \nonumber \\
&&  \left.s_{23}^2-2 c_{12}^2 c_{\rho } \left(2 s_{23}^2-1\right)
s_{\delta +\rho } y_1^2-2 c_{12}^2 c_{\rho } s_{\delta +\rho }
y_3^2\right)- \left.\left.2 c_{12}^2 c_{23}^2 c_{\delta +\rho }
s_{23}^2 s_{\rho }
\left(y_2^2-y_3^2\right)\right]\right\} \nonumber \\
&+&\frac{\zeta _{32}}{2 c_{12} c_{23} s_{12} s_{13} s_{23}}
\left\{c_{12} c_{23} s_{12} s_{13} s_{23} \left[\left(\left(2 s_{2
\sigma }+s_{2 \delta +2 \sigma }\right) \left(y_3^2-y_2^2\right)
s_{23}^2 +s_{2 \sigma } \left( y_2^2- y_3^2\right)
\right.\right.\right. \nonumber \\ && \left.
+\left(s_{13}^2+1\right) s_{2 \delta +2 \sigma } y_1^2+s_{13}^2 s_{2
\delta +2 \sigma } \left(y_3^2-y_2^2\right) s_{23}^2 -s_{13}^2 s_{2
\delta +2 \sigma } y_3^2-s_{2 \delta +2 \sigma } y_3^2\right)
\nonumber \\ && \left.s_{12}^2+\left(2 s_{23}^2-1\right) s_{2 \sigma
} \left(y_2^2-y_3^2\right)-s_{13}^2 s_{2 \delta +2 \sigma }
\left(\left(y_3^2-y_2^2\right) s_{23}^2+y_1^2-y_3^2\right)\right]
\nonumber \\ && -  c_{12}^2 \left[\left(s_{13}^2
\left(\left(s_{\delta }+3 s_{\delta +2 \sigma }\right)
\left(y_2^2-y_3^2\right) s_{23}^4+\left(2 \left(s_{\delta }+2
s_{\delta +2 \sigma }\right) y_3^2-2 s_{\delta +2 \sigma }
y_2^2\right) s_{23}^2 \right.\right.\right. \nonumber \\ && -
\left.2 c_{\sigma } \left(2 s_{23}^2-1\right) s_{\delta +\sigma }
y_1^2-2 c_{\sigma } s_{\delta +\sigma } y_3^2\right) \nonumber \\
&& \left.\left.\left. - 2 c_{23}^2 c_{\delta +\sigma } s_{23}^2
s_{\sigma } \left(y_2^2-y_3^2\right)\right) s_{12}^2+2 c_{23}^2
c_{\delta +\sigma } s_{13}^2 s_{23}^2 s_{\sigma }
\left(y_2^2-y_3^2\right)\right]\right\} \nonumber \\ &+&\frac{1}{2
c_{12} c_{23} s_{12} \zeta _{21}}\left\{-2 \left(c_{\delta +\rho
-\sigma }-2 c_{\delta } c_{\rho -\sigma } s_{12}^2\right) s_{13}
s_{23} s_{\rho -\sigma } \left(y_2^2-y_3^2\right) c_{23}^2 \right.
\nonumber \\ && -\left.c_{12} s_{12} s_{2 \rho -2 \sigma }
\left[-c_{13}^2 y_1^2-\left(s_{13}^2 s_{23}^2-c_{23}^2\right)
y_2^2+\left(s_{23}^2-c_{12}^2 s_{23}^2\right)
y_3^2\right]c_{23}\right\} \nonumber \\
&+&\frac{1}{2 c_{12} c_{23} s_{13} s_{23} \zeta _{31}} \left\{c_{12}
c_{23} s_{13} s_{23} \left[\left(-\left(2 s_{2 \rho }+s_{2 \delta +2
\rho }\right) \left(y_2^2-y_3^2\right) s_{23}^2
\right.\right.\right. \nonumber \\ &&  \left.+s_{13}^2 s_{2 \delta
+2 \rho } \left(y_3^2-y_2^2\right) s_{23}^2+ s_{2 \rho } y_2^2-2
c_{\delta } s_{\delta +2 \rho } y_3^2-s_{13}^2 s_{2 \delta +2 \rho }
y_3^2\right) s_{12}^2 \nonumber \\ && + \left.\left(s_{12}^2
\left(s_{13}^2+1\right)-1\right) s_{2 \delta +2 \rho } y_1^2+s_{2
\delta +2 \rho } \left(\left(y_2^2-y_3^2\right)
s_{23}^2+y_3^2\right)\right] \nonumber \\ && + s_{12}
\left[\left(-\left(\left(s_{\delta }-3 s_{\delta +2 \rho }\right)
s_{12}^2+4 c_{\delta +\rho } s_{\rho }\right)
\left(y_2^2-y_3^2\right) s_{23}^4 \right.\right. \nonumber \\ && +
\left(2 \left(c_{\delta +\rho } s_{\rho }-s_{12}^2 s_{\delta +2 \rho
}\right) y_2^2-2 \left(\left(s_{\delta }-2 s_{\delta +2 \rho
}\right) s_{12}^2+3 c_{\delta +\rho } s_{\rho }\right) y_3^2\right)
s_{23}^2  \nonumber \\ && + \left.2 c_{12}^2 c_{\delta +\rho }
\left(2 s_{23}^2-1\right) s_{\rho } y_1^2+2 c_{12}^2 c_{\delta +\rho
} s_{\rho } y_3^2\right)  \left.\left.s_{13}^2+2 c_{12}^2 c_{23}^2
c_{\rho } s_{23}^2
s_{\delta +\rho } \left(y_2^2-y_3^2\right)\right]\right\} \nonumber \\
&+&\frac{1}{2 c_{12} c_{23} s_{12} s_{13} s_{23} \zeta _{32}}
\left\{\left[\left(s_{13}^2 \left(-\left(s_{\delta }-3 s_{\delta +2
\sigma }\right) \left(y_2^2-y_3^2\right) s_{23}^4
\right.\right.\right.\right. \nonumber \\ && -2 \left(s_{\delta +2
\sigma } y_2^2+\left(s_{\delta }-2 s_{\delta +2 \sigma }\right)
y_3^2\right) s_{23}^2- \left.2 c_{\delta +\sigma } \left(2
s_{23}^2-1\right) s_{\sigma }
y_1^2-2 c_{\delta +\sigma } s_{\sigma } y_3^2\right) \nonumber \\
&& - \left.\left.\left.2 c_{23}^2 c_{\sigma } s_{23}^2 s_{\delta
+\sigma } \left(y_2^2-y_3^2\right)\right) s_{12}^2+2 c_{23}^2
c_{\sigma } s_{13}^2 s_{23}^2 s_{\delta +\sigma }
\left(y_2^2-y_3^2\right)\right)\right]c_{12}^2  \nonumber \\ && +
c_{23} s_{12} s_{13} s_{23} \left[\left(s_{13}^2 s_{2 \delta +2
\sigma } \left(y_2^2-y_3^2\right) s_{23}^2+\left(2 s_{2 \sigma
}+s_{2 \delta +2 \sigma }\right) \left(y_2^2-y_3^2\right) s_{23}^2
\right.\right. \nonumber \\ && - \left.\left(s_{13}^2+1\right) s_{2
\delta +2 \sigma } y_1^2-s_{2 \sigma } y_2^2+s_{2 \sigma }
y_3^2+s_{13}^2 s_{2 \delta +2 \sigma } y_3^2+s_{2 \delta +2 \sigma }
y_3^2\right) \nonumber \\ && \left.\left.s_{12}^2-\left(2
s_{23}^2-1\right) s_{2 \sigma } \left(y_2^2-y_3^2\right)+s_{13}^2
s_{2 \delta +2 \sigma } \left(\left(y_3^2-y_2^2\right)
s_{23}^2+y_1^2-y_3^2\right)\right]c_{12}\right\} \, .
\end{eqnarray}
In addition, the analytical RGEs for $\rho$ and $\sigma$ can be
obtained by combining Eqs.~\eqref{eq:RGEkappa}, \eqref{eq:U}, and
\eqref{eq:para}.  However, the corresponding formulas are rather
lengthy, and therefore, we do not list these tedious results here,
since the evolution behavior is well described by
Eqs.~\eqref{eq:phase-approx1}-\eqref{eq:phase-approx3}.

\bibliography{bib}
\bibliographystyle{apsrevM}

\end{document}